\documentclass[linenumbers,twocolumn]{aastex631}
\usepackage{amsmath, systeme, amssymb}
\usepackage{bm}% bold math
\usepackage{csquotes}
\usepackage{graphicx}
\usepackage[utf8]{inputenc}
\usepackage[english]{babel}
\usepackage[T1]{fontenc}
\usepackage{enumerate}
\usepackage{longtable}
\usepackage{supertabular}
\usepackage{lineno}
\usepackage[normalem]{ulem}
\usepackage{hyperref}
\hypersetup{
    colorlinks,
    citecolor=blue,
    filecolor=blue,
    linkcolor=blue,
    urlcolor=blue
} 
\usepackage[normalem]{ulem}
\usepackage{bbm}
\usepackage{xcolor}
\begin{document}

\nolinenumbers

\title{Boosting decision trees for Main Belt Asteroid selection in planetary ephemerides: an alternative model \footnote{Released on XXX}}

\author{Vincenzo Mariani}
\affiliation{Université Côte d'Azur, Observatoire de la Côte d'Azur, CNRS}
\affiliation{CRAS, Sapienza University of Rome, via Eudossiana 18, 00184 Rome, Italy}

\author{Agnès Fienga}
\affiliation{Université Côte d'Azur, Observatoire de la Côte d'Azur, CNRS}

\author{Zachary Murray}
\affiliation{Université Côte d'Azur, Observatoire de la Côte d'Azur, CNRS}

\author{Mickaël Gastineau}
\affiliation{LTE, Observatoire de Paris, Université PSL, Sorbonne Université, Univ. Lille, Laboratoire National de Métrologie et d'Essai, CNRS, 75014 Paris, France}

\author{Jacques Laskar}
\affiliation{LTE, Observatoire de Paris, Université PSL, Sorbonne Université, Univ. Lille, Laboratoire National de Métrologie et d'Essai, CNRS, 75014 Paris, France}

%\collaboration{20}{(AAS Journals Data Editors)}

%\author{F.X Timmes}
%\affiliation{Arizona State University}
%\affiliation{AAS Journals Associate Editor-in-Chief}

%\altaffiliation{AASTeX v6+ programmer}

%\affiliation{AAS Director of Publishing}
%\affiliation{American Astronomical Society \\
%1667 K Street NW, Suite 800 \\



\begin{abstract}

\nolinenumbers
One of the main bottleneck in assessing the accuracy of Mars orbit is the unknown value of the asteroids in the Main Asteroid Belt. Nowadays a modeling with 343 asteroids as point masses is used, with the relative masses fitted to observational data. In the current work we propose an innovative methodology to reduce the number of asteroids implemented as point masses, thus reducing the number of parameters to be fitted, without a significant degradation of the postfit residuals.

\end{abstract}


\section{Introduction} \label{sec:intro}

%\subsection{Planetary ephemerides}\label{sec:intro:PE}

To elaborate an accurate prediction of the position of a celestial body of the solar system is a problem rooted in the beginning of the astronomy. Over the time, the quantity of observations increased and the quality improved. 
%Before 1960s the only available observations for planetary positions were optical (measures of angles). Together with the first observations of the distance between celestial bodies and the Earth (radar measurements for Venus, laser for the Moon, Viking mission for Mars), the orbits accuracy has been largely improved with respect to the ones obtained with just optical observations \citep{Newhall_al_1983}. The development of numerical solutions of planetary dynamics, then, became essential to reach observational accuracy in the orbits computations. Since the 1970s the first numerically integrated planetary ephemerides were built \citep{standish1976jpl}. 
So far, several tools for planetary orbitography have been developed from different teams around the world. Among them, the INPOP planetary ephemerides \citep{inpop21a} have been developed since 2003, and they are based on the numerical integration of a dynamical model for the planetary orbits including several effects: of the relativistic $N$-body problem considering the eight planets, Pluto, Moon and asteroids, TNOs \citep{FiengaMNRAS2019, inpop21a}, {as well as} Sun oblateness and Lense-Thirring effect for what concerns solar physics \citep{Fienga2015}. {Moreover the Earth-Moon interaction is considered, taking into account the Moon interior \citep{viswa2017PhD}.} 
%Nowadays, modern planetary orbitography is done by integrating numerically the Einstein-Infeld-Hoffmann equations of motions as proposed by \cite{moyer1971mathematical, Moyer2000}, and by adjusting the parameters of the dynamical model to solar system observations such as space navigation ans radio science data, ground-based optical observations and lunar laser ranging \citep{Fienga2008, Fienga2011, inpop21a}. 
In order to fit the constants of the dynamical model the least squares method is used \citep{book:TapleySchutzBorn}, applying the Boundary Value Least Squared (BVLS) method \citep{Lawson1995} to minimize the weighted sum of the residuals. The metric used to assess the goodness of the fit, i.e. what we minimize with the BVLS procedure, is the $\chi^2$ that we define as

\begin{equation}\label{eq:chi2_from_dyn_sys_P}
\chi^2 (P) = \frac{1}{N_{\text{obs}}} \sum_{j=1}^{N_{\text{obs}}} \left( \frac{g^j(P)  - d^j_{\text{obs}} }{\sigma_j} \right)^2   . 
\end{equation}

where $P$ are the parameters fitted, $\bm{d}_{\text{obs}}=(d_{\text{obs}}^j)_j$ is the set of observations, $g^j$ represents the computation of the observables corresponding to $d_{\text{obs}}^j$ and $\sigma_j$ are the observational uncertainties.
The $\chi^2$ defined with Eq. \eqref{eq:chi2_from_dyn_sys_P} depends upon some parameters $P$, therefore we have the function
\[ P \longmapsto \chi^2(P) \] that associates to the values of $P$ the corresponding value of $\chi^2(P)$. 
In Eq. \eqref{eq:chi2_from_dyn_sys_P} we neglected the dependencies on time for sake of simplicity, however the reader should keep in mind that they are taken into account during the $\chi^2$ computation. \\

%\subsection{Main Asteroid Belt perturbations}\label{sec:Intro:MAB_pert}

The question of modeling the Main Asteroid Belt (MAB) perturbations on the inner planet orbits has been addressed since 1980s \citep{WILLIAMS19841} {with the goal of finding a reliable model, for many thousands of objects with unknown masses, which may have a significant gravitational influence with the inner planets (mainly Mars).} \citet{WILLIAMS19841} proposed a specific list of 343 individual objects among the Main Belt Asteroids (MBAs), gathering the most perturbing objects to consider for the construction of accurate planetary ephemerides. 

%In particular, for modern-day planetary orbitography 343 perturbing asteroids as point masses are considered, with their individual masses fitted. The list of the most perturbing asteroids is based upon an analytical study  of the commensurability among the orbital periods of Mars and of the asteroids \citep{WILLIAMS19841, Standish1998_AA}. 
For the heaviest asteroids, such as Ceres and Vesta, the relativistic contribution is considered next to the Newtonian one \citep{inpop21a}. For the smallest asteroids, instead, only the Newtonian part is accounted in the equations of motion. The orbits of the MBAs are integrated together with the orbits of the planets, within a full $N$-body problem framework, although the mutual interactions among asteroids are not accounted, except for the 5 biggest (Ceres, Pallas, Vesta, Hygieia and Interamnia).

%The four innermost planets are the most influenced by the MBAs {Jupiter? Or is it where we have the best constraints?}. 

In the case of Mars, the range measurements on the Earth-Mars distance have an accuracy of about 1~m, whereas the influence of the MBAs on the Mars orbit can reach a few kilometers \citep{StandishFienga2002}. If we desire to predict the position of Mars with an uncertainty comparable to the observational accuracy, the asteroid perturbations have to be estimated extremely accurately, therefore potentially we need a model at $0.1 \%$ or even less. {The orbits of the asteroids are usually known, but their masses are not, except for specific cases.  Therefore most of the uncertainty is related to the mass, more than to the orbit \citep{CARRY201298, FiengaMNRAS2019, Mariani2024PhD}.}

{Different strategies have been proposed to overcome the problem of chosing the most appropriate selection of asteroids perturbing the planets (i.e. \cite{StandishHellings1989,StandishFienga2002, SOMENZI2010858, Kuchynka2010,KUCHYNKA2013243,FiengaMNRAS2019})}.

In the current work we describe an innovative methodology {which we use} to obtain a ranking, by relative importance, of the 343 asteroids currently used in the MAB modeling for planetary orbitography \citep{inpop21a}. Starting from this ranking, we provide a reduced list of asteroids {whose absence} do not degrade significantly the accuracy of the fit, reducing the number of components of the vector $P$ (of the overall fit) by more than {140 components}. 
To reach this aim, first we use a supervised learning tool called Boosting Decision Trees \citep{Breiman1984_book, ESLII_Hastie} to obtain the ranking. We {then} check the goodness of the ranking fitting for the asteroid masses to the full observational dataset using the MAB modeling with a reduced number of asteroids. {The ranking plays a key role since it allows to test which asteroids might be removed and which ones are relevant, following their relative importance {in terms of gravitational perturbations}. Full results are shown in Sec. \ref{sec:meth}.}  

\section{Boosting decision tree for asteroid ranking}\label{sec:meth}

%Machine learning and data-driven approaches are becoming more and more important in many areas of science. 
The aim of machine learning is to make predictions from training examples. {These techniques have now become} {an important} part of the algorithmic toolbox of a researcher. The goal of this section is to describe the tools of machine learning we used to tackle the problem of asteroid selection in the Main Belt: the boosting decision trees (BDT).
Among the several machine learning methods, gradient tree boosting has the big advantage {of providing} a ranking of the variables in input to the regression function. This aspect, next to increasing the interpretability of the regression itself, plays a key role in our work to obtain a ranking by relative importance of the masses for the {343 asteroids of the MAB modeling as described beforehand.} 
%\cafo{For this work, we used the package \texttt{XGBoost} \citep{xgboost2016}.}
{In this section, we give only the main steps of the method. The full algorithm and explanation are given in \citep{Mariani2024PhD}}.

%\af{TO BE REDUCED =============== dont forget to use \citep{Mariani2024PhD} as a reference}

% ==============================================

\subsection{Methods: Boosting decision trees} 
Understanding the gravitational influence of asteroids in the Main Belt is crucial for high-precision solar system dynamics, particularly in spacecraft navigation and {ephemeris} modeling. This study leverages Boosted Decision Trees (BDT) to rank the 343 asteroids based on their impact on the residuals of Mars Express (MEX) tracking data. The goal is to identify which asteroids significantly perturb the dynamical model, allowing for targeted refinements in planetary ephemerides.  

\subsubsection{ Boosted Decision Trees (BDT) for Regression} 
Decision Trees (DT) are a supervised learning method \citep{ESLII_Hastie} that partitions the domain into rectangular regions and fits a constant value to each. Such a partition can be represented, by construction, as a binary {decision} tree {(BDT)}. While a single DT provides a piecewise-constant approximation on each region \citep{Breiman1984_book, ESLII_Hastie}, combining multiple regression trees via boosting \citep{Friedman2001, ESLII_Hastie} (sequentially added regression decision trees) improves predictive accuracy.  
Each new tree focuses on the residuals (errors) of the previous tree prediction with respect to the values of the training set, refining the ensemble of trees iteratively. Each DT is built by splitting along one of the axes in the $n$-dimensional domain of the variables in input for the function one wants to reproduce. Each new rectagle of the partition, corresponds to a split along an axis, as well as to a split in the binary tree representing the partition. The algorithm used for the training \citep{xgboost2016} ranks the input variables (here, asteroid masses) by: i) the frequency of splits for each input variable, i.e. how often a variable is used to partition the data; ii) the improvement in fit induced by that split, characterized by the reduction in prediction error attributed to each new partition region. The investigation uses XGBoost \citep{xgboost2016}, an optimized gradient-boosting framework that incorporates regularization to prevent overfitting (the reader is addressed to \cite{xgboost2016} and references therein).  

\subsubsection{Defining the Target Function}  \label{sec:target_function_def}
The strategy consists in reproducing a  target function $f^* = f^*(\Delta \bm{m}^{\text{AST}})$, function of the asteroid masses, using a BDT. As a consequence, the ranking of the masses $\bm{m}^{\text{AST}}$ is obtained, based on the relative importance for each one of the components of $\bm{m}^{\text{AST}}$. Indeed, thanks to the BDT, it is possible to know the relative importance for each component of $\bm{m}^{\text{AST}}$ in reproducing $f^*$.
The $f^*$ we chose quantifies how changes in asteroid masses $ \Delta \bm{m}^{\text{AST}} $ affect the goodness-of-fit metric $\chi^2$, as defined in Eq. \eqref{eq:chi2_from_dyn_sys_P} , for MEX tracking residuals. We built, then, a training set to be provided to the BDT algorithm to reproduce $f^*$. 
For sake of computational {efficiency (see the discussion in \cite{Mariani2024PhD}}, we approximated $\chi^2$ with a quadratic expression, resulting in a target $f^*$ as in the following equation:
\begin{equation}\label{eq:sec2:target_function_appr}
f^*: \Delta \bm{m}^{\text{AST}} \mapsto \Delta \tilde{\chi}^2_{\mathcal{O}_{\text{MEX}}}(\Delta \bm{m}^{\text{AST}})    
\end{equation}

In Eq. \eqref{eq:sec2:target_function_appr} $\Delta \tilde{\chi}^2$ is the approximate change in $\chi^2$ (a measure of fit quality) induced by a change in the masses $\Delta \bm{m}^{\text{AST}}$, whereas $ \mathcal{O}_{\text{MEX}}$  denotes the hypothesis of restricting $\chi^2$ to the MEX spacecraft observations.  

\subsubsection{Training Set Construction}\label{sec:training_set_construction}

The training set we need to generate, then, is 
\begin{equation}\label{eq:sec2:tr_set_nom}
    \mathcal{D} = \{ (\Delta \bm{m}_i, \Delta \tilde{\chi}^2_{i, \mathcal{O}_{\text{MEX}}}) \}_{i=1}^M
\end{equation} %\Delta \tilde{\chi}^2_{\mathcal{O}_{\text{MEX}}}(\Delta \bm{m}^{\text{AST}})_i)
where in Eq. \eqref{eq:sec2:tr_set_nom} $\Delta \tilde{\chi}^2_{i, \mathcal{O}_{\text{MEX}}} = \Delta \tilde{\chi}^2_{\mathcal{O}_{\text{MEX}}}(\Delta \bm{m}^{\text{AST}}_i))$ and $M$ is the size of the training set. For what concerns the mass variations $(\Delta \bm{m}_i)_i$, they have been generated randomly with a uniform distributions within $\pm 10 \% $ of the postfit values from the INPOP21a ephemerides model \citep{inpop21a}. Then, for each perturbation $\Delta \bm{m}_i = \Delta \bm{m}$, we compute the corresponding $\Delta \tilde{\chi}^2_i$ using linearized residuals as follows:
\begin{equation}\label{eq:sec2:chi2tilde}
\Delta \Tilde{\chi}^2_{\mathcal{\mathcal{O}}} (\Delta \bm{m}) \equiv \left[ \Delta \widetilde{(O-C)} \rvert_{\mathcal{\mathcal{O}}} \right]^T \left[ \Delta \widetilde{(O-C)} \rvert_{\mathcal{\mathcal{O}}} \right] .
\end{equation}
%       \Delta \tilde{\chi}^2_i = \left[ \widehat{\Delta(O-C)} \right]^T \left[ \widehat{\Delta(O-C)} \right],  
where $\mathcal{O} = \mathcal{O}_{\text{MEX}}$ and
\begin{equation}\label{eq:sec2:omc_tilde}
%    \Delta \widetilde{(O-C)} \rvert_{\mathcal{O}} = \Delta \widetilde{(O-C)} \rvert_{\mathcal{O}} (\Delta \bm{m}) \equiv \left[ \frac{\partial (O-C) \rvert_{\mathcal{O}} }{\partial \bm{m}} \right] \Delta \bm{m}.
\Delta \widetilde{(O-C)} \rvert_{\mathcal{O}} \equiv \left[ \frac{\partial (O-C) \rvert_{\mathcal{O}} }{\partial \bm{m}} \right] \Delta \bm{m}.
\end{equation}

%\begin{equation}\label{eq:linear_oc_def}
%    \Delta \widetilde{(O-C)} \rvert_{\mathcal{\mathcal{O}}} =  
%\end{equation}
is the estimated change in MEX tracking residuals. The linearity expressed in Eq. \eqref{eq:sec2:omc_tilde} is an assumption, as well as the restriction to the MEX data. The validity of such assumptions is going to be proved by the results {presented in Sect. \ref{sec:ranking_fit}}. Finally, then, the training set \( \mathcal{D} = \{ (\Delta \bm{m}_i, \Delta \tilde{\chi}^2_i) \}_{i=1}^M \) links mass changes to their dynamical impact.  

%\subsection{Asteroid Ranking}

The BDT model processes $\mathcal{D}$ to assign an importance score to each asteroid, reflecting its influence on MEX residuals. 
%We then obtained a subset of asteroids (e.g., Ceres, Vesta) likely dominates due to their large masses.
{We also see that the importance is relative to other asteroids and to the specific datasets used ($\mathcal{O}_{\text{MEX}}$, i.e. ranging from Mars Express).} 
{The limitation of this approach is mainly in the linearized residual of Eq. \eqref{eq:sec2:omc_tilde} approximation that may miss nonlinear effects; future work might consider full numerical simulations for the production of the training set. }

\subsection{Planetary ephemerides improvement}

\begin{figure}
 % 283 & $3.01 \times 10^{18}$ & $9.89 \times 10^{17}$ & $2.64 \times 10^{18}$ & 0.3284 & 0.37 \\ 
%    \includegraphics[scale=0.35]{Figures/Pie3_lab_4E4.pdf}
    \includegraphics[scale=0.2]{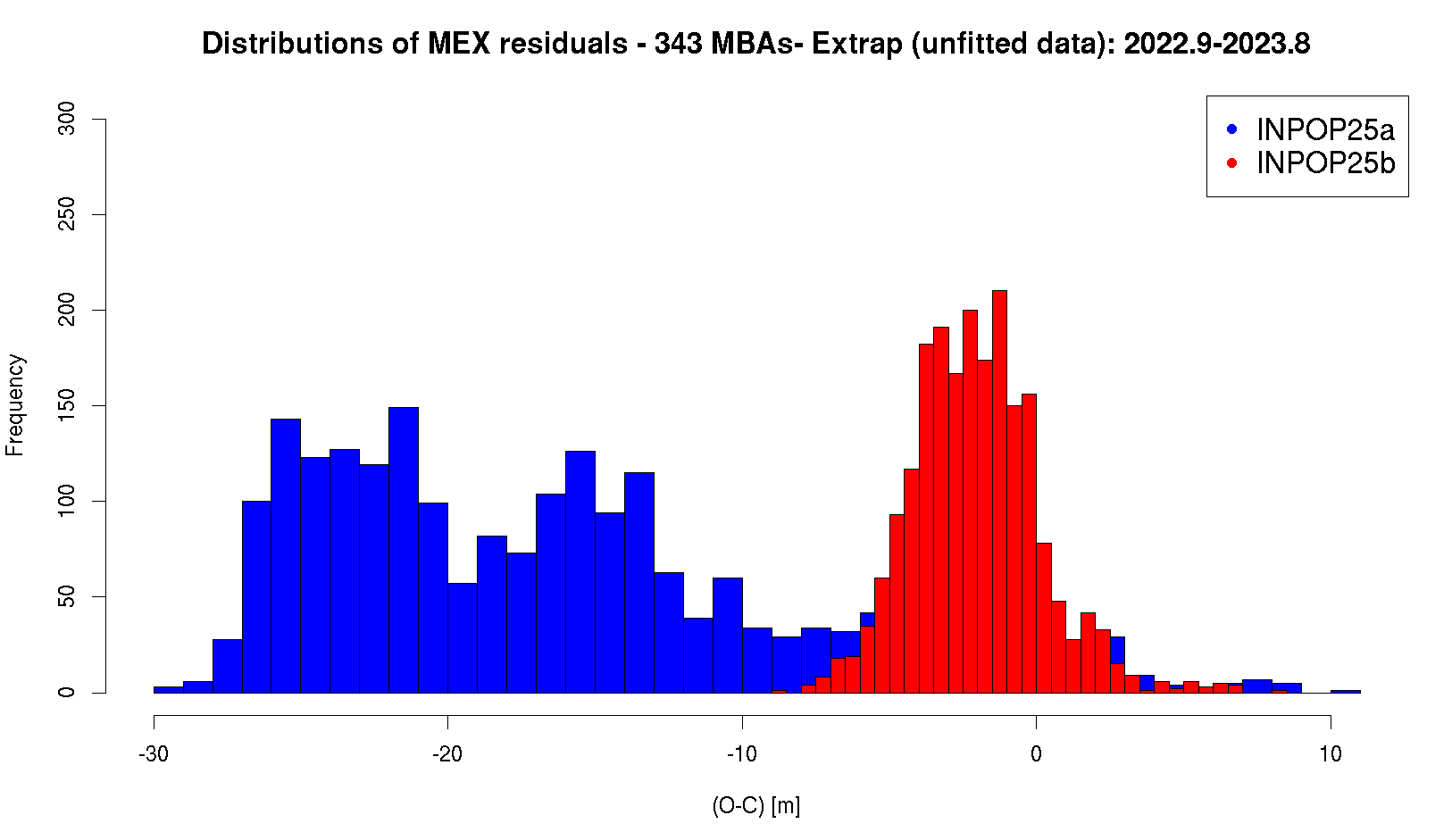}
    \caption{Distributions of residuals for the extrapolation period using the original boundary interval (in blue) and the extended boundaries (in red).}
    \label{fig:extraphist}
\end{figure}

For implementing the ranking established with the {BDT}, we had started from an updated version of INPOP planetary ephemerides, that we will call INPOP25a.
This version is built using the same procedure as INPOP21a \cite{inpop21a} but the interval of adjustment has been prolonged, from 2019 to 2022. As it was already noticed in \citep{FiengaMNRAS2019}, the current limitation of planetary ephemerides is mainly their capability of extrapolation for the Mars geocentric distances. 
%It is a issue that has already addressed in the literature (see i.e \cite{KUCHYNKA2013243, FiengaMNRAS2019}) and that is clearly visible on Fig. \ref{fig:extraphist}, where residuals obtained after the the period of adjustment are  plotted. 
In \cite{FiengaMNRAS2019}, it has been proposed that the most up-to-date knowledge of asteroid bulk densities should taken into account using random Monte Carlo sampling of the boundaries used for planetary ephemeris construction. INPOP19a, INPOP21a and the most recent INPOP25a have been built using such a procedure, in keeping the same boundaries issued from the \cite{FiengaMNRAS2019}.
However, as it is clearly visible on Fig. \ref{fig:extraphist}, the issue of the extrapolation is still present. So, to improve, it was decided to relax the boundary conditions by 25$\%$, as a way to limit the over-constraint imposed on the adjustment which might explain the bad extrapolation capability of the model. With the new obtained ephemeris fitted on the same data sample as INPOP25a, we obtained the results presented on the bottom plot of Fig. \ref{fig:extraphist} which indicates a clear improvement of the extrapolation of the new solution (INPOP25b). The improvement affects not only the dispersion of the residuals but also their general distributions: the new INPOP25b producing close to normal distribution whereas INPOP25a distribution seems to be composed by several multiple distributions, indicating remaining signals. A small offset is still present but can be explained by the MEX transponder delay that is not fitted during the computation of extrapolated residuals. As a consequence, for the rest of this study, we implement the new asteroid ranking starting from the INPOP25b version. 
%}

%\begin{figure}
%    \centering
%    \includegraphics[scale=0.5]{Figures/extrap21a25ab.png}
%    \caption{MEX postfit and extrapolated one-way residuals in meters computed with INPOP21a, INPOP25a and the INPOP25a with extended boundaries (INPOP25b). The solid vertical lines indicate the limit of the adjustement periods and the beginning of the extrapolation one.}
%    \label{fig:extrapres}
%\end{figure}
%\end{figure}

\subsection{Ranking}\label{sec:2:ranking}

\begin{figure}
    \centering
    \includegraphics[width=0.5\textwidth]{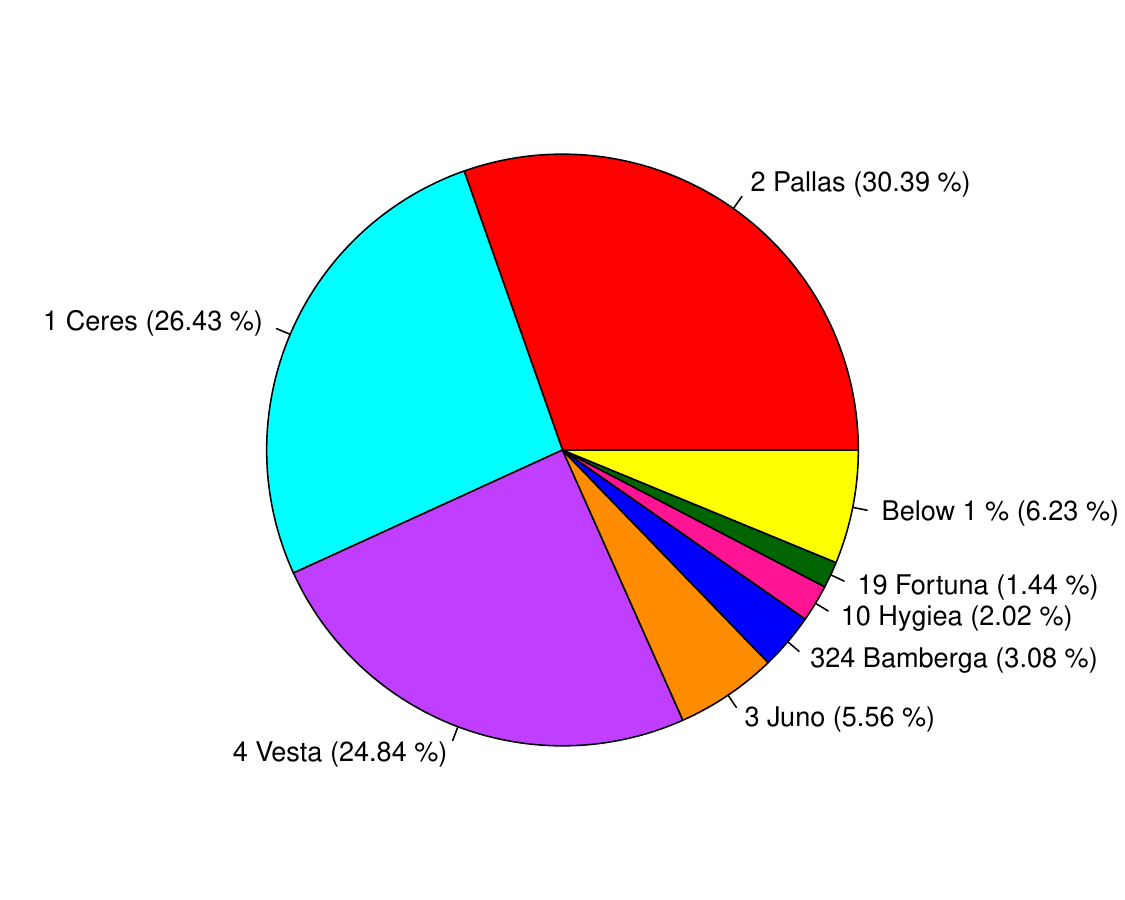}
    \caption{{Pie chart of the relative importance for the first asteroids in the ranking obtained with $2 \times 10^6$ elements.}}
    \label{fig:Pie_2E6_depth17}
\end{figure}

As explained in Secs. \ref{sec:training_set_construction} and \ref{sec:target_function_def}, the idea is to use a BDT for the regression of a training set $\mathcal{D}$, as defined in Eq. \eqref{eq:sec2:tr_set_nom}, from which we obtain a ranking of the variables in input to the function $f^*$ of Eq. \eqref{eq:sec2:target_function_appr}. As a byproduct, we get the relative importance (and the ranking) of the components in $\Delta \bm{m}^{\text{\tiny AST}}$, representing the asteroid masses. Within our study we produced several training sets, all of them define through Eqs. \eqref{eq:sec2:tr_set_nom} and \eqref{eq:sec2:chi2tilde}, using different sizes (i.e. different $M$ in Eq. \eqref{eq:sec2:tr_set_nom}) and different hyperparameters configurations for the trees. We present {in Fig. \ref{fig:Pie_2E6_depth17}} only the outcome of the solution we got with $M = 2 \times 10^6$, being the best result we have found so far. The relative variable importance is provided in a way such that the sum of them, for all the asteroids, is 1. Therefore, it is easy to interpret the importance for each asteroid in percentage with respect to the global one. %\cafo{In Fig. \ref{fig:Pie_2E6_depth17} we see the outcome, with a pie chart for the relative importance with $M = 2 \times 10^6$.} 
As we can see {in Fig. \ref{fig:Pie_2E6_depth17}}, more than $80 \%$ of the global importance is taken from just three asteroids: 1 Ceres, 2 Pallas and 4 Vesta {with importances between $20 \%$ and $30 \%$}. Following them, with an importance between $1 \%$ and $5 \%$ we find 3 Juno, 324 Bamberga, 10 Hygiea and 19 Fortuna. Subsequently, only asteroids with a relative importance below $1\%$ have been found. From this first computation it already turns out how the majority of the asteroids have small relative importance alone, but not negligible importance when considered all together. This first outcome also confirms the very widespread correlation among the asteroids in the planetary orbitography fit, as already pointed out in \cite{KUCHYNKA2013243, FiengaMNRAS2019} and \cite{Mariani2024PhD}. In other words, the effect of the ensemble of asteroids is not negligible. In order to have a clear assessment of the ranking obtained, however, we need to use it with the full fit of planetary ephemerides. In Sec. \ref{sec:ranking_fit} we see how to introduce the ranking in the fit to improve the dynamical model. 

\subsection{Introduction of the ranking in the fit}\label{sec:ranking_fit}

\begin{figure}
    \centering
    \includegraphics[scale=0.45]{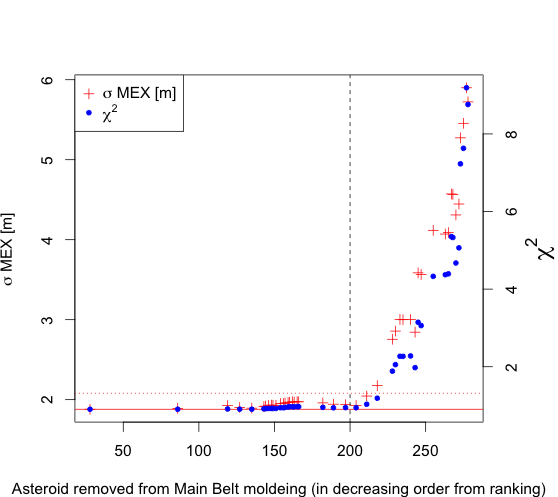}
    \caption{Standard deviation ($\sigma_{\text{MEX}}$) of the MEX residuals in meters and global $\chi^2$ {obtained with}
    %between the reference modeling (i.e. INPOP25a) and%
    the model with $X$ asteroids removed. $X$ is given by the x-axis according to the ranking given by Sec. \ref{sec:2:ranking}. The dotted line represent a reference of 20~cm {increase relative to INPOP25b $\sigma_{\text{MEX}}$ plotted with the red full line.}.}
    \label{fig:Sigmas_Mex_PostfitChi2_Postfit}
\end{figure}

The ranking obtained with the BDT provides a relevant information about the importance of each asteroids in the global reconstruction of the Earth-Mars distances as deduced from the MEX navigation data. The point is, then, to see how effective is this ranking when one wants to consider the global construction of the whole planet orbitography. To do so, we started from one initial planetary ephemeris and we removed one by one the asteroids following the BDT ranking, starting from the less important objects. By removing the asteroid, one means that we do not consider the object in the dynamical modeling anymore and consequently we do not estimate its mass in the planetary global fit. We built a full ephemeris in removing the considered objects and in fitting iteratively the obtained integration. For the fit, we stop considering only the MEX tracking data but we operate the adjustment of the 8 planets of the solar system using the complete set of available observations. For this study, we used the 
%\cafo{INPOP21a datasets \citep{inpop21a}}
{INPOP25a datasets}. The removal of the asteroids is done by accumulation, meaning that for the first object removed, the integration counts 342 asteroids and when the asteroid ranked 200 in the BDT list is removed, 143 asteroids perturb the planet orbits. Convergency is obtained before the $9^{\text{th}}$ iteration and the results of the fitted ephemerides  in terms of standard deviations of the MEX residuals $\sigma_{\text{MEX}}$ and in the global $\chi^2$ are shown in Fig. \ref{fig:Sigmas_Mex_PostfitChi2_Postfit}.

%\begin{figure}
%    \centering
%    \includegraphics[scale=0.22]{Figures/Plot_sigmas_2E6d17_075.pdf}\\\%includegraphics[scale=0.22]{Figures/Plot_Chi2_2E6d17_075.pdf}
%    \caption{\vm{Differences in the standard deviation (Top Panel) %($\sigma_{MEX}$) of the MEX residuals in meters and global $\chi^2$ %(Bottom Panel) between the reference modeling (i.e. INPOP) and the %model with $X$ asteroids removed. $X$ is given by the x-axis according %to the ranking given by Sec. \ref{sec:rank}. The dashed line represent %a reference of 20~cm}.}
%    \label{fig:Sigmas_Mex_PostfitChi2_Postfit}
%\end{figure}

%\begin{figure}
%    \centering
%    \includegraphics[scale=0.35]{Figures/Plot_Chi2_2E6d17_075.pdf}
%    \caption{Differences in the global $\chi^2$ between the %reference modeling (i.e. \vm{INPOP25b}) and the model with $X$ %asteroids removed. $X$ is given by the x-axis according to the %ranking obtained in each one of the cases.}
%    \label{fig:Chi2_Postfit}
%\end{figure}

With Fig. \ref{fig:Sigmas_Mex_PostfitChi2_Postfit}, one can draw several conclusions. The evolution of $\sigma_{\text{MEX}}$ and of the global $\chi^2$ follow very similar trends, meaning that indeed MEX is a good proxy of the quality of the global ephemerides. The ranking is validated by the increase of the differences to the reference solution {(INPOP25b)} with the increase of the ranking. 
%\cafo{The differences in $\sigma_{\text{MEX}}$  present however some oscillations that do not compromise the general trend. These oscillations indicate some noise in the asteroid ranking consistent the sensitivity of the ranking to the dynamical modeling.}
In Fig. \ref{fig:Sigmas_Mex_PostfitChi2_Postfit} mainly two regimes can be observed: the first asteroids removed, so presenting the lowest importance according to BDT, induced indeed very small differences {(below 20~cm)} in $\sigma_{\text{MEX}}$ and the global $\chi^2$. After some limit (up to rank 200), the differences start to increase more, almost with an exponential trend. From Fig. \ref{fig:Sigmas_Mex_PostfitChi2_Postfit} one can see how removing up to 200 asteroids yields to an oscillations in the post-fit $\sigma_{\text{MEX}}$ less than 20~cm.

\subsubsection{The residual indicator and selection}\label{sec:res_indic}

\begin{figure*}
    \centering
    \includegraphics[scale=0.5]{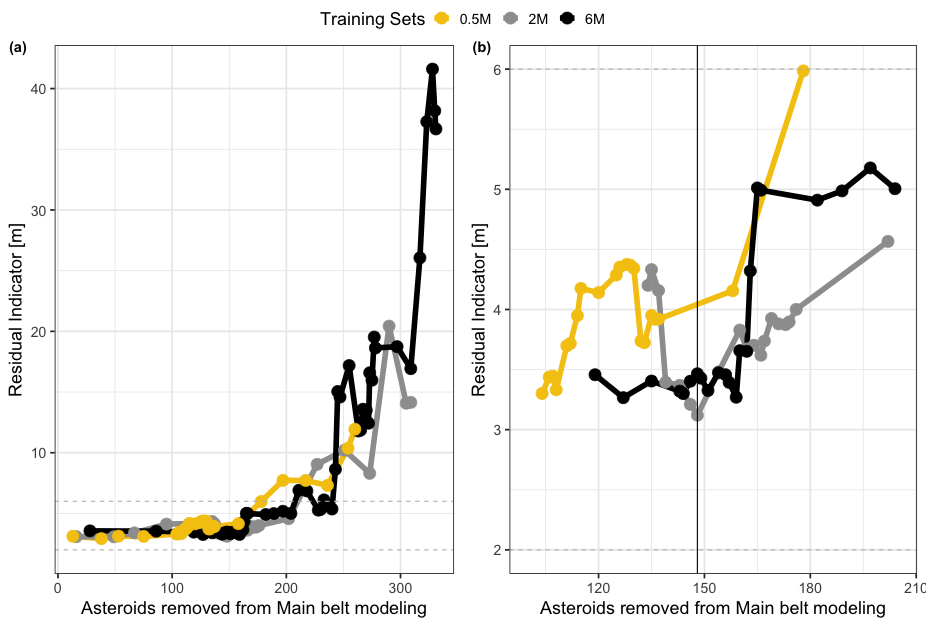}
    \caption{{Residual Indicator} obtained taking into account the $\sigma_{\text{MEX,fit}}$ obtained after fit and the $\sigma_{\text{MEX,ext}}$ obtained on the extrapolation interval for different sizes of the training sets. Panel (b) is an enlargement of Panel (a). The grey vertical line indicates the minimum reached by the residual indicator $R_I$. The corresponding solution also maximize the number of asteroids removed from the Main belt original selection. {The different curves correspond to results obtained starting from training sets of different sizes as indicated by  the legend on the top (0.5 million, 2 millions and 6 millions).}}
    \label{fig:Average_Index_Postfit}
\end{figure*}

In order to propose a new solution of planetary ephemerides with {an optimized number of perturbing MBAs}  modeled in the main belt, but without degrading the postfit residuals, we can take into account next to $\sigma_{\text{MEX}}$ and $\chi^2$, also the extrapolation capability of the solution found. 
In order to do so we computed, for the several cases we analysed, an averaged index that we call residual indicator $R_I$. 
The $R_I$ takes into account both the $\sigma_{\text{MEX}}$ on the observations used in the fit ($\sigma_{\text{MEX,fit}}$) and the dispersion of the residuals from extrapolation of MEX unfitted data ($\sigma_{\text{MEX,ext}}$). 
The RI is defined as {$R_I = \sqrt{\left( \sigma_{\text{MEX,fit}}^2 + \sigma_{\text{MEX,ext}}^2 \right)}$} and it is computed removing the asteroids in a cumulative way, starting from the less important. 
Computing the $R_I$ for different training sets $\mathcal{D}$ allows to assess which solution is the best compromise among amount of asteroids removed, postfit $\sigma_{\text{MEX}}$ and extrapolation capability on unfitted data. 
Looking for the minimum of the $R_I$ provides what we consider being the {\it best solution}, and we call it INPOP25c. 
The results are shown in Fig. \ref{fig:Average_Index_Postfit}. In removing more than 100 asteroids we can propose a solution at 147 asteroids removed, obtained from the ranking with $M=2\times10^6$ elements. 
In the left panel of Fig. \ref{fig:Average_Index_Postfit} three different sizes of training sets are analysed, with $M=5\times10^{5}$, $M=2\times10^{6}$ or $M=6\times10^{6}$. In the right panel of Fig. \ref{fig:Average_Index_Postfit} an enlargement of the left panel is presented on the best solution found.  

\section{Discussion and conclusion}\label{sec:discussion}

So far we have shown that we can approximately reproduce and even improve upon the extrapolation of INPOP25a and INPOP25b with INPOP25c and significantly fewer asteroids, especially when the bounds on the masses are relaxed. This naturally raises the question about wether these new masses are physically reasonable.% or if they are forced to extreme values in order to fit the data.

\subsection{Comparisons between INPOP versions}
\label{sec:comp-in}
\begin{figure}
    \centering
    \includegraphics[scale=0.5]{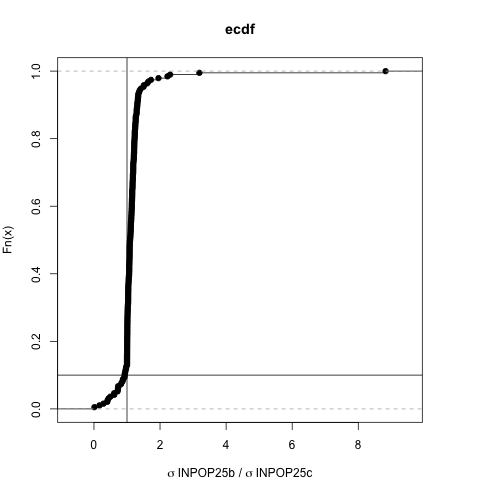}
    \caption{Cumulative histogram of the ratio $\sigma_{25b}$ over $\sigma_{25c}$ for the 193 asteroid masses fitted in both INPOP25b and INPOP26c.}
    \label{fig:ecdf_sig}
\end{figure}

In order to assess the advantages of reducing the number of asteroids in the point of view of asteroid mass determination, we compare the uncertainties on fitted masses obtained with INPOP25b ($\sigma_{25b}$) and with INPOP25c ($\sigma_{25c}$), for the common masses. In Fig. \ref{fig:ecdf_sig} is plotted the cumulative histogram (empirical cumulative distribution function) of the ratio between $\sigma_{25b}$ and $\sigma_{25c}$ for the 196 masses fitted both by INPOP25b and INPOP25c. As it is clearly visible, 90$\%$ of uncertainties obtained with INPOP25b are larger than the one obtained with INPOP25c. In average, the improvement induced by the reduction of the number of fitted asteroids in INPOP25c is of about 15$\%$, compared to INPOP25b.
This improvement can be explained by a better conditioning of the Jacobian matrix of the fit with a reduction of 70$\%$ of the conditioning number between INPOP25b and INPOP25c. The decrease of the conditioning number is likely due to the smaller set of asteroids taken into account during the fit of  INPOP25c with respect to INPOP25b. In terms of individual improvements we count 10 masses with a decrease of the mass uncertainties better than 50$\%$, and 32, better than 25$\%$. All the masses together with their uncertainties are given in Appendix (Table \ref{tab:appendix1}).
So, the improvement brought by the reduction of the number of asteroids accounted for in the fit is clearly visible in terms of mass uncertainties and quality of the fit. Furthermore, the choice of INPOP25c was also made in order to have a good trade-off between postfit residuals and extrapolation as discussed in Sec. \ref{sec:res_indic}. Finally it is interesting to note that by having less asteroids in the dynamical modeling, we {reduce} the time of computation of about 52$\%$ (usertime) and 30$\%$ (realtime) in comparison with INPOP25b integration.

\begin{figure}    
%    \centering
    \includegraphics[scale=0.3]{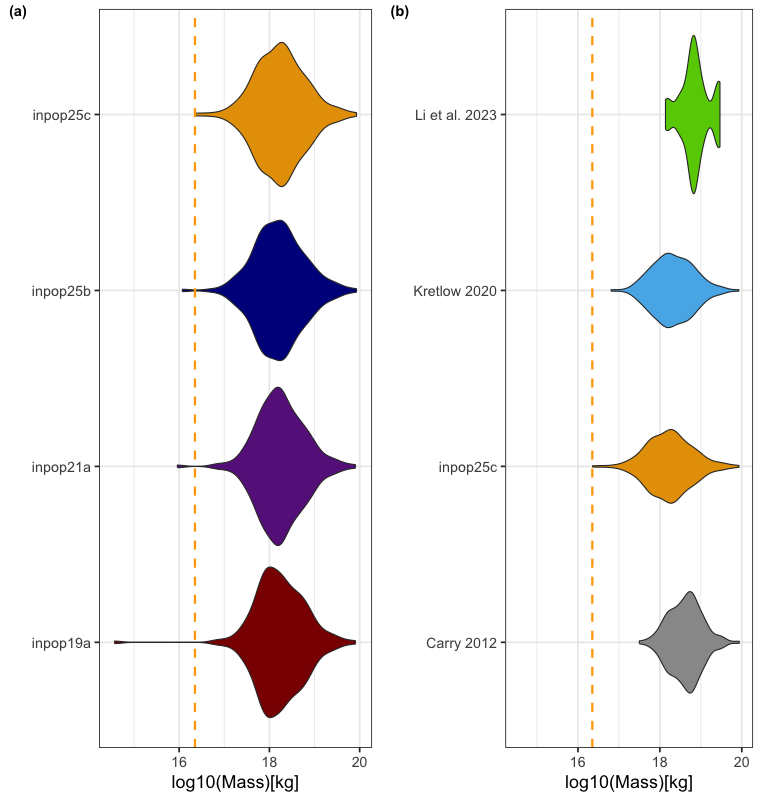}
    \caption{Histograms of the masses of asteroids common to INPOP25b, INPOP26c, INPOP21a \citep{inpop21a},  INPOP19a \citep{INPOP19a_Fienga}, \cite{Li_2023}, \cite{Kretlow2020} and \cite{CARRY201298}.}
    \label{fig:violon_mass2}
\end{figure}

{More globally, we compare the posterior distributions of masses obtained with INPOP19a \citep{INPOP19a_Fienga}, INPOP21a \citep{inpop21a}, INPOP25b and INPOP25c, presented on the Panel a of Fig. \ref{fig:violon_mass2}. As one can see on these plots, the INPOP mass distributions are quite similar from one ephemerides to another. It is interesting to note, however, that INPOP19a tends to give smaller masses and with ephemerides involved with more data and progress in the modeling, the dispersion of the mass distribution reduces. Differences on mass distributions are more important between other determinations as explained in Sect. \ref{sec:other_comp}.}
%\cvm{For the conditioning number, see \cite{LawsonHanson1995}.}

%\begin{figure}
%    \centering
%    \includegraphics[scale=0.35]{Figures/Plot_Index_4E4_2E6d17_v1.pdf}
%    \includegraphics[scale=0.35]{Figures/ecdf_inpop.png}
%    \caption{Cumulative histograms of the masses of asteroids common to INPOP25b, INPOP26c, INPOP21a and INPOP19a. In red, cumulative histogram of the ratio $\sigma_{25b}$ over $\sigma_{25c}$ for the 193 asteroid masses fitted in both INPOP25b and INPOP26c. \af{If the ratio is greater then one (dashed vertical red line), this means that the uncertainties of the INPOP25b masses are greater than the INPOP26c ones.}}
%    \label{fig:ecdf_inpop}
%\end{figure}

\subsection{Comparison with independent mass estimations}

\subsubsection{Neural-network albedo algorithm}\label{sec:NN_albedo}

To assess if the masses obtained  with INPOP25c are physically reasonable, we require an independent posterior of asteroid masses.  We generate this posterior by using the neural-network catalog of albedo predictions presented in \cite{Murray_2023}. This approach uses an ensemble of neural networks to predict an asteroids visible albedo - and the uncertainty thereof, as a function of it's proper orbital elements. Given the predicted albedo from that catalog ($p$) and the total uncertainty ($\sigma_p = \sqrt{\sigma_{belt}^2 + \sigma_{pred}^2}$) we can generate a distribution for the albedo for our objects. We note that four cases MPC (132),(433),(454) and (1036), no proper elements were available. In these cases the albedo was assumed to be like in a log-uniform distribution between 0.01 and 1.0.  These distributions in albedo were then converted into distributions of diameters using H magnitudes provided by the Minor Planet Center. Finally, a posterior distribution in mass was obtained by assuming the density was a uniform distribution between 0.5 - 4.5~$\mathrm{ g/cm^{3}}$.

To test whether our values are extreme we consider the distribution of the percentile scores of each of our fit masses with respect to the corresponding posterior distributions.  If the generated posterior distributions are the real distributions, the distribution of the posterior scores should be a uniform distribution.  If the fit masses are too constrained (or the distributions too wide), they'll be disproportionally found near the 50th percentile.  In the opposite case, they'll be found disproportionnally found at 0th or 100th percentile.

\begin{figure}
    \centering
    \includegraphics[scale=0.35]{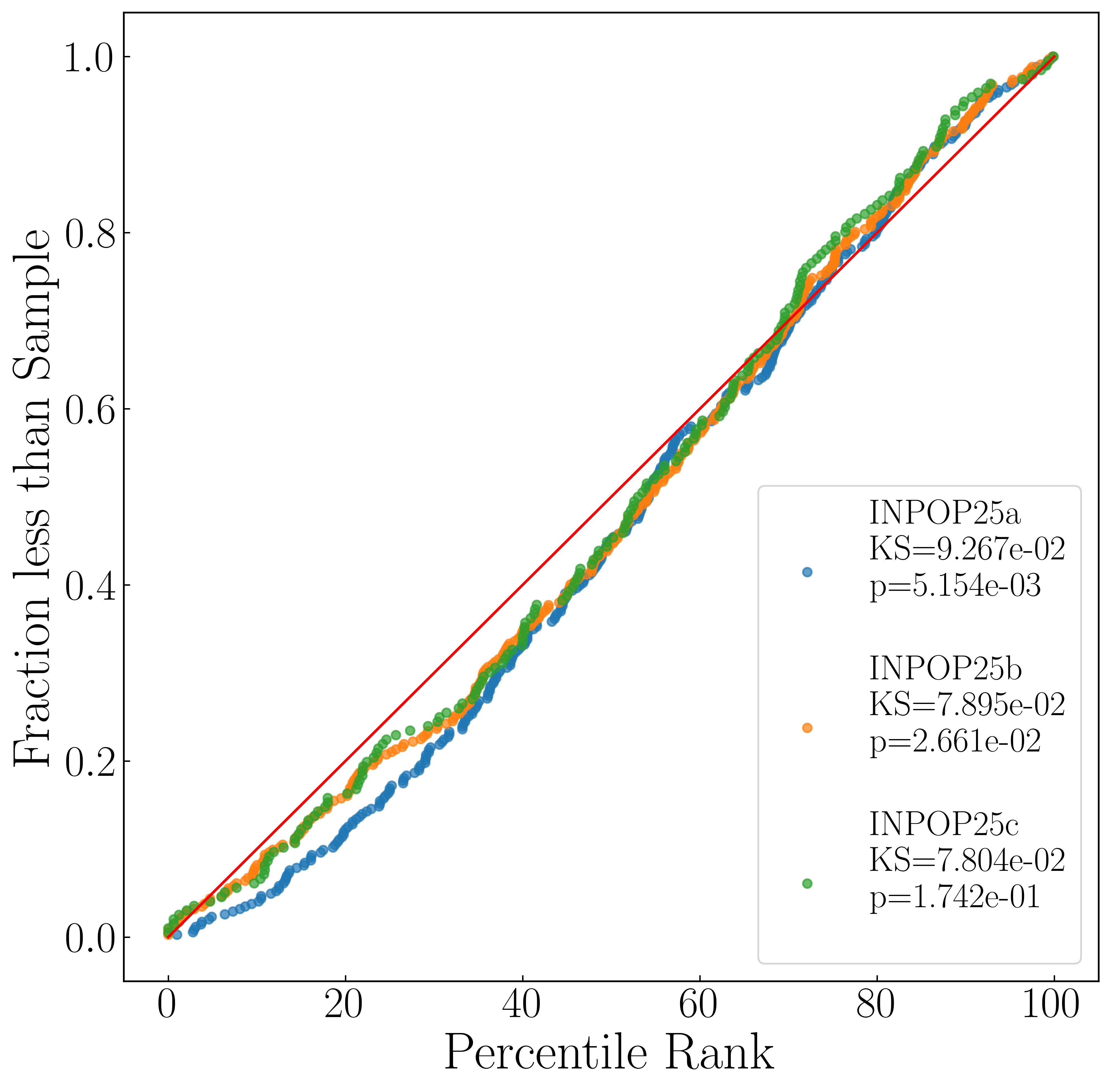}
    \caption{Here we show a comparison of the ECDF of the distribution of percentile scores of our fitted masses with respect to the posterior mass distributions.  Perfect agreement is denoted by the red diagonal line.  We see the distributions found in INPOP25c are comparable with those found in INPOP25a and INPOP25b.  We also provide p values for our tests, to assess the statistical significance of our comparisons.}
    \label{fig:ecdf}
\end{figure}

To test a set of asteroids against each other we can simply preform a Kolmogorov-Smirnov test on the resulting percentile distribution and check its proximity to a uniform distribution.  The closer the proximity, the more likely it is that the fit masses are consistent with the posterior distribution.  We show a summary of the ECDF of these distributions in {Fig.} \ref{fig:ecdf}. We see that while all of our distributions are slightly centrally concentrated - likely owing to the wide distributions of densities we consider, there is little difference between the selection of asteroids in this work and those in previous INPOP versions. This suggests that the masses we find are not physically unreasonable compared to those studies. 

%A full 0.5
%B full 0.75
%C 145 0.75

%\begin{figure}
%    \centering
%    \includegraphics[scale=0.4]{Figures/ecdf_compar.png}
 %   \caption{Cumulative Histograms of masses from INPOP25c, \cite{CARRY201298}, \cite{Li_2023} and \cite{Kretlow2020}.}
%    \label{fig:ecdfcomp}
%\end{figure}

\subsubsection{Comparisons with literature}
\label{sec:other_comp}

We have compared masses issued from the metadata analysis from \cite{CARRY201298} and the recent update by \cite{Kretlow2020} and from close-encounters analysis using Gaia data by \cite{Li_2023}.  The complexity of providing consistent and homogeneous database for masses comes from the difficulty of associating mass determinations obtained with different methods (planetary ephemerides, binary system, asteroid-asteroid close-encounters), more or less impacted by different sources of uncertainties and biases.  On one hand, \cite{CARRY201298} addressed this issues in using an iterative weighted estimation of the mass, rejecting outliers by comparison to the mean and $\sigma$ values, considering also estimations without published uncertainties. On the other hand,  \cite{Kretlow2020} have added supplementary mass determinations compared to \cite{CARRY201298} and had used more sophisticated statistical estimators  (EVM) that should be less sensitive to outliers and provide asymmetric uncertainties \citep{BIRCH2014106}. In \cite{Li_2023}, are reported 20 masses that have been obtained by close-encounters observed with GAIA DR3 and ground-based observations.

%\begin{itemize}
%    \item for global comparisons, without considering the common masses:\\

\begin{figure*}
    \centering
    \includegraphics[scale=0.35]{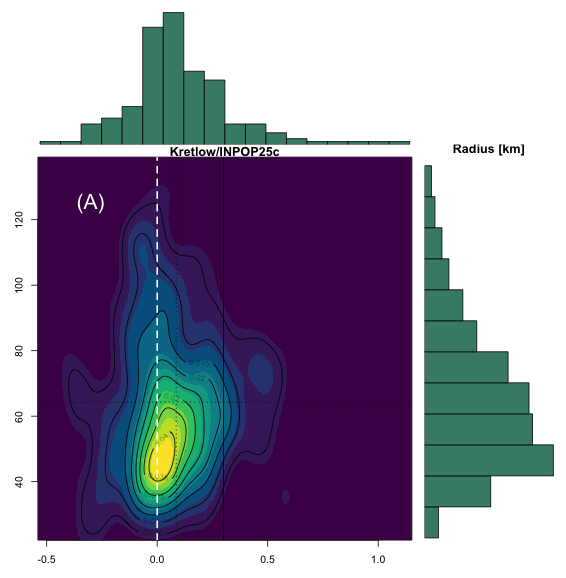}\includegraphics[scale=0.35]{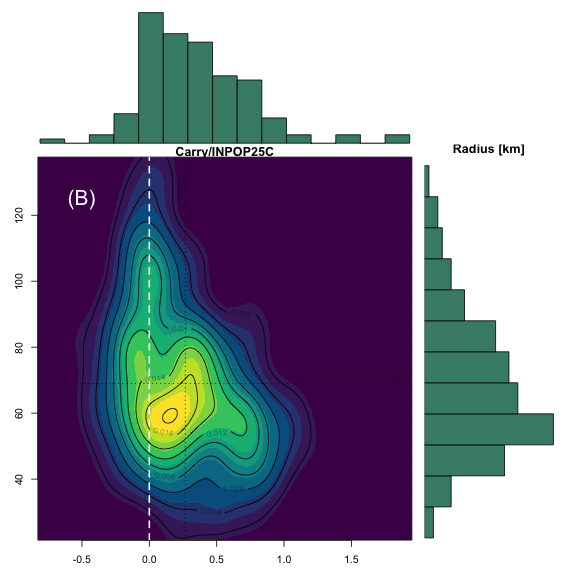}
        \includegraphics[scale=0.35]{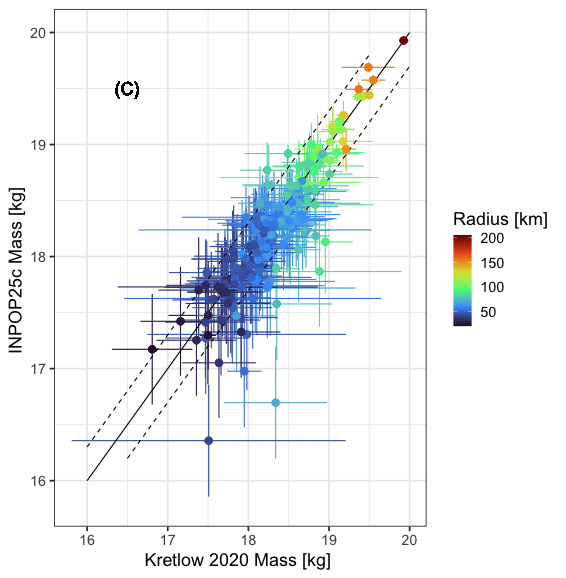}
    \includegraphics[scale=0.35]{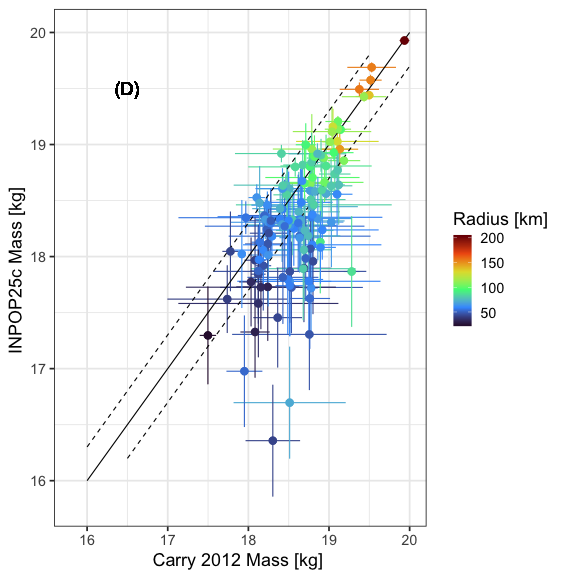}
    \caption{Comparisons between INPOP25c, \cite{CARRY201298} and \cite{Kretlow2020}. Panels A and B give the 2D histograms of the radii in kilometers and ratio \cite{Kretlow2020}/INPOP25c (A)  and the ratio \cite{CARRY201298}/INPOP25c (B). On Panels C and D, one finds the masses of INPOP26c versus the  \cite{Kretlow2020} ones (C) and masses of INPOP26c versus \cite{CARRY201298} (D). The color indicates the radii in kilometers.}
    \label{fig:ratio_radius}
\end{figure*}

A first global comparison is possible with the Panel b of Fig. \ref{fig:violon_mass2}. On this panel, are plotted the mass distributions common to INPOP25c, \cite{CARRY201298},\cite{Kretlow2020} and  \cite{Li_2023}. It appears clearly that INPOP25c and  \cite{Kretlow2020} have very close distributions with the same number of objects obtained with both distributions. \cite{CARRY201298} seems to indicate higher masses than the two former but with 30$\%$ less objects.  \cite{Li_2023} also gives higher masses than \cite{Kretlow2020} and INPOP25c, but this could also be related to the method (close-encounters favouring heavier objects) and the limited number of masses (21) provided by \cite{Li_2023}. 
The fact that the comparison of the global distribution is affected by the distribution of the asteroid sizes is also visible in Fig. \ref{fig:ratio_radius} with the Panels A and B. Where we show 2D histograms of the radii (in y-axis) and of the ratio between the \cite{Kretlow2020}  and INPOP25c masses (Panel A) and between the \cite{CARRY201298}  and INPOP25c masses (Panel {B}). The mean of the radii of the common asteroids between  \cite{Kretlow2020}  and INPOP25c is of about 64~km but it is about 8$\%$ bigger between \cite{CARRY201298}  and INPOP25c. 
Furthermore,  one can also see Fig. \ref{fig:ratio_radius}, that for asteroids with radii greater than 70~km, the masses are pretty consistent, with ratio close to 1 on Panels C and D or close to 0 in log10 scale on Panels A and B. 

However, for smaller objects, the discrepency in masses increases. For the \cite{Kretlow2020} versus INPOP25c comparison (Panels A and C of Fig. \ref{fig:ratio_radius}), the departure from unity is balanced between under and over-estimated INPOP25c masses, with a small bias toward under-estimated INPOP25c. One can also noticed a slight departure for asteroids bigger than 100~km radius for which the \cite{Kretlow2020} masses seem to be smaller than INPOP25c. However, this dispersion can be explained by high uncertainties from \cite{Kretlow2020} determinations (see Panel C of Fig.\ref{fig:ratio_radius}). 
For the \cite{CARRY201298} versus INPOP25c comparison (Panels B and D of Fig. \ref{fig:ratio_radius}), there is a clear bias with \cite{CARRY201298} being systematically bigger than INPOP25c for asteroids with radii smaller than 70~km. But contrary to the comparison with  \cite{Kretlow2020}, this bias between \cite{CARRY201298} and INPOP25c can not be explained by uncertainties as {shown in} Panel D.
So in conclusion, one can see that, as expected INPOP25C, \cite{Kretlow2020} and \cite{CARRY201298}  have some systematics differences,  but one can also note a better agreement between INPOP25C and \cite{Kretlow2020}. 

%\end{itemize}
%\begin{figure*}
%    \centering
%    \includegraphics[scale=0.4]{Figures/dm2jL}
%    \caption{Comparisons between INPOP25c and \cite{CARRY201298} masses  versus radius (Panel (a) with squares for \cite{CARRY201298}) and between INPOP26c and \cite{Kretlow2020} masses versus radius (Panel (b) with triangles for  \cite{Kretlow2020}).The color indicates the Noise over Signal ratio ($N/S = \sigma_{mass}/{\textrm{mass}}$).}
%    \label{fig:dmcomp}
%\end{figure*} 

%\begin{figure*}
%    \centering
%    \includegraphics[scale=0.15]{Figures/mass_mass}
%    \caption{Comparisons between INPOP25c and \cite{CARRY201298} masses  versus radius (Panel (b) ) and between INPOP26c and \cite{Kretlow2020} masses versus radius (Panel (a)).The color indicates the radii in kilometers.}
%    \label{fig:dmcomp}
%\end{figure*}

%\begin{figure*}
%    \centering
%   \includegraphics[scale=0.3]{Figures/h2D_kretlow25c.png}\includegraphics[scale=0.3]{Figures/h2D_carry25c.png}
%    \caption{\af{2D histograms of the radii in kilometers and ratio between the masses obtained with \cite{CARRY201298} and INPOP25c (left-hand side) and between the one obtained with \cite{Kretlow2020} and INPOP25c (right-hand side) }}
%    \label{fig:ratio_radius}
%\end{figure*} 

%\begin{figure*}
%    \centering
%    \includegraphics[scale=0.15]{Figures/mass_mass}\includegraphics[scale=0.3]{Figures/h2D_kretlow25c.png}
%    \includegraphics[scale=0.15]{Figures/mass_mass}\includegraphics[scale=0.3]{Figures/h2D_carry25c.png}
%    \caption{\af{2D histograms of the radii in kilometers and ratio between the masses obtained with \cite{CARRY201298} and INPOP25c (left-hand side) and between %the one obtained with \cite{Kretlow2020} and INPOP25c (right-hand side) }}
%    \label{fig:ratio_radius}
%\end{figure*} 

\subsection{Combining determinations: Wasserstein tools}

\begin{figure}    
    \centering
    \includegraphics[scale=0.4]{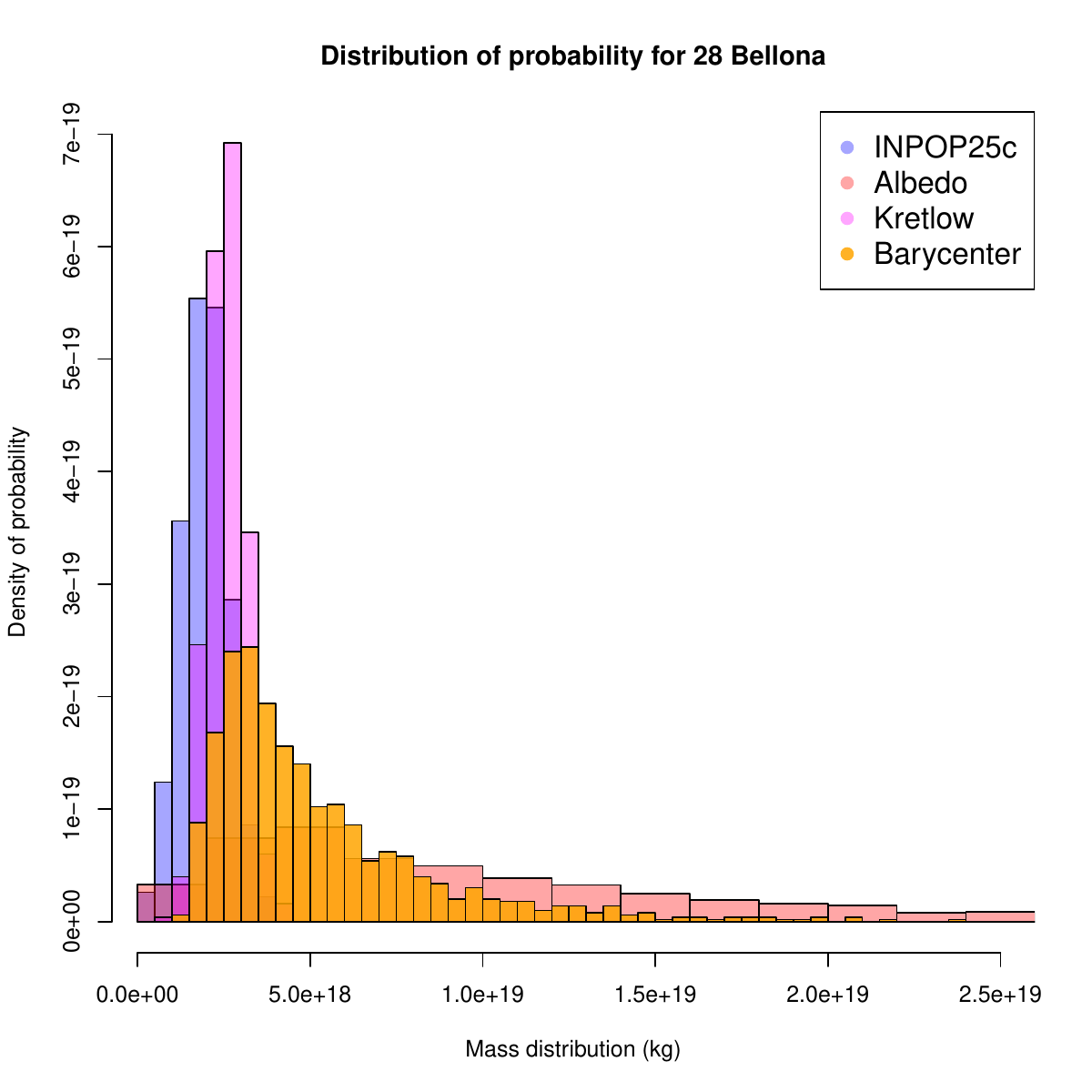}
    \caption{Histograms of the mass distributions for 28 Bellona estimated from INPOP25c, neural networks and albedo (Sec. \ref{sec:NN_albedo} and \cite{Murray_2023}) and \cite{Kretlow2020}. In yellow the Wasserstein barycenter resulting from the three posteriors cited.}
    \label{fig:28Bellona}
\end{figure}

\begin{figure}    
    \centering
    \includegraphics[scale=0.25]{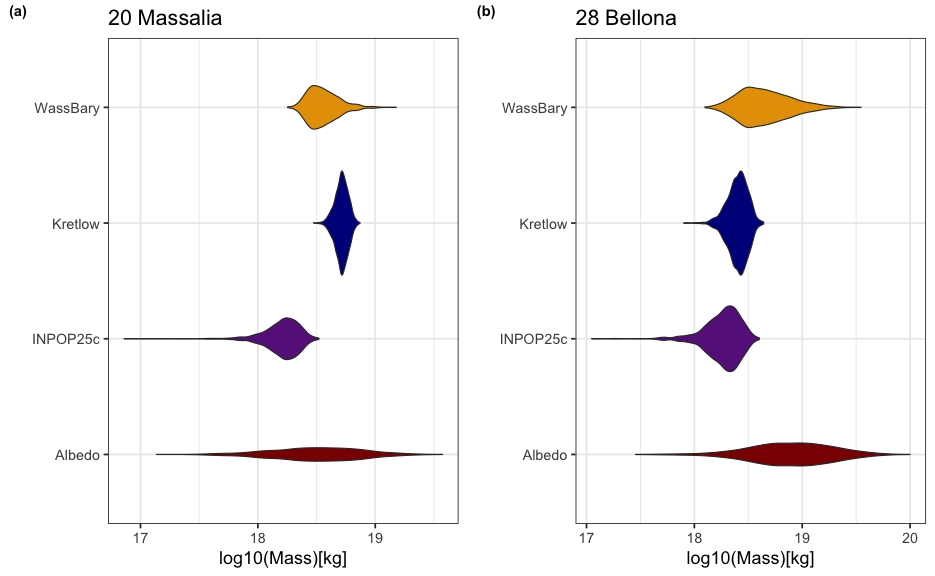}
    \caption{Histograms of the mass distributions for 28 Bellona estimated from INPOP25c, neural networks and albedo (Sec. \ref{sec:NN_albedo} and \cite{Murray_2023}) and \cite{Kretlow2020}. In yellow the Wasserstein barycenter resulting from the three posteriors cited.}
    \label{fig:28Bellona_violins}
\end{figure}

{Finally, we performed a direct comparison between the probability distribution for the value of each MBA found with this methodology \citep{Murray_2023}, and the postfit masses obtained in the case of INPOP25c, i.e. the optimal {planetary} solution with fewer asteroids. 
%\sout{From INPOP25c we took both the postfit masses found and their relative uncertainties ($\sigma_{25c}$). Starting from these, we produce synthetically a normal distribution for each one of the asteroids, having as mean value the postfit mass and as standard deviation the uncertainty found with the BVLS algorithm (see Sec. \ref{sec:intro}).} 
%\zm{The INPOP25c distributions were produced assuming a normal distribution with mean and standard deviation ($\sigma_{25c}$) \vm{as postfit values} given by the BVLS algorithm (see Sec. \ref{sec:intro}).}
{To compare} these distributions {with those from other sources} {we propose} an innovative approach from the theory of optimal transport and calculus of variation \citep{villani2008optimal, book_FigalliGaludo2023}, {specifically} the computational adaptation of the Wasserstein barycenters by means of the package \texttt{waspr} \citep{WassBary_R_pack2020} based on the work of \cite{PUCCETTI2020104581}. 
%\zm{Not sure what you're saying here. Do you mean:..computation of Wasserstein barycenters by means of the package...? ..computation of Wasserstein barycenters and mean using the package...? Or something else?}
{The Wasserstein barycenter is a method to \textit{average} two (or more) generic  probability distributions 
%\caf{we should find a way to say this idea differently}
in the space of probability distributions, {which we apply to the asteroid mass distributions}. The Wasserstein barycenter, in particular, does not require for the distributions involved to have the same support (like in the case of the KL divergence), or a specific law explicit in analytical formulation. Therefore it can be applied as far as one consider the same physical quantity in the same unit.} The underlying idea is to find a way  {to merge} the information contained in several different posterior distributions available for each asteroid considered in INPOP25c, {and to join them} {using} a consistent probabilistic approach, {and without making any hypothesis about the law followed by the distribution}.  {This merging is especially important for small asteroids, whose mass estimates can differ significantly between studies.} {Consequently,} the Wasserstein barycenter\footnote{The reader should note that we use this tool since it allows a fully consistent comparison among probability distributions of whatever shape and support.} % (differently from the KL-divergence, where the support must be the same when you want to proceed with a such a computation). 
is, then, not a single value but a probability distribution. For a wider introduction to the topic of computational optimal transport the reader is addressed to \cite{peyre2020compOT_book} and references therein. The interest in using the Wasserstein barycenters plays a role within the context of the comparison shown in Sec. \ref{sec:other_comp}. 
{As already pointed out in Sect. \ref{sec:comp-in} and \ref{sec:other_comp},} for smaller asteroids the difference in their mass estimations are larger among the different studies. 
Using the barycenters, we propose a way to join the information (including their uncertainties) and methodologies of the previous investigations together with the mass determinations found during the fit procedure of INPOP25c.
In Appendix, Table \ref{tab:appendix1} we report the masses found with INPOP25c, as well as the mean of the Wasserstein barycenters. 
%\sout{For any given asteroid which mass is fitted in INPOP25c,} 
The Wasserstein barycenters are computed between 
%\sout{the normal distribution for the possible values of the mass, from postfit mean and standard deviation}
{the normal distributions} obtained with INPOP25c, the posterior ditributions obtained using albedos and neural networks \citep{Murray_2023}, as well as with the normal distributions proposed by \cite{Kretlow2020}.
We show here some specifc examples to {illustrate} the use of the Wasserstein barycenter, in particular in Fig. \ref{fig:20Massalia} we show {the case of the mass of} (20) Massalia, and in Fig. \ref{fig:28Bellona} we show {the one of} (28) Bellona. In the case of 20 Massalia, we see from Fig. \ref{fig:20Massalia} that the estimations from INPOP25c and \cite{Kretlow2020} are both with small uncertainty ($\sigma$), 
%accurate in terms of standard deviations \zm{what does `accurate in terms of standard deviations' mean?  Do you mean with 3 sigma of each other? Do you mean the uncertainties are small? Or something else?}, 
and within the uncertainty of the albedo estimation, however, there is a discrepancy among the two normal distributions. In such a case the Wasserstein barycenter  provides an intermediate estimation among the three cases.
\begin{figure}    
%    \centering
    \includegraphics[scale=0.4]{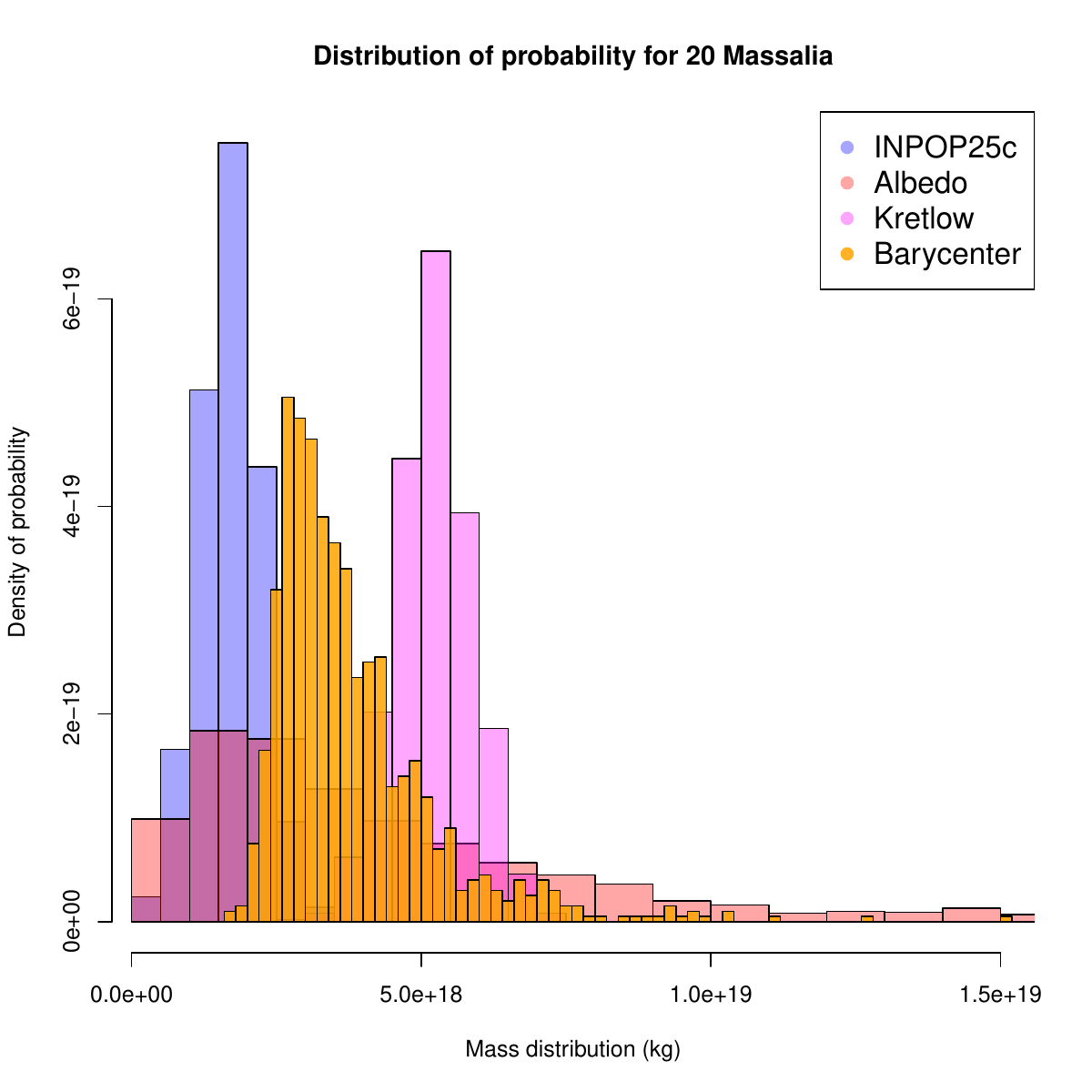}
    \caption{Histograms of the mass distributions for 20 Massalia estimated from INPOP25c, neural networks and albedo (Sec. \ref{sec:NN_albedo} and \cite{Murray_2023}) and \cite{Kretlow2020}. In yellow the Wasserstein barycenter resulting from the three posteriors cited.}
    \label{fig:20Massalia}
\end{figure}
Also, in the case of 28 Bellona, we see an example of Wasserstein barycenter that is not a normal distribution but more similar to a log-normal, showing the flexibility and generality of the tool that might handle different kind of uncertainties and distributions all at the same time. {Furthermore, in Fig. \ref{fig:28Bellona_violins} we can see the several distributions involved for the computation of the barycenter -- both for 20 Massalia, panel (a) and 28 Bellona, panel (b) -- visualized as disentangled distributions to stress their differences even for the same asteroid.} %
%\caf{where is the rest of the sentence ?}
We point out, however, that the mean and standard deviation obtained from the computation of the barycenters is not to be intended as a better estimation of what has been found with the previous mass determinations. On the other hand it is a tool that might be used for specific cases of asteroids for which there is poor information and discrepancy among the different estimations, becoming a way to unify them in a probabilistic approach.}

\section{Conclusions}

{We have presented our work on the use of boosting decision trees in order to get a ranking by relative importance of the point mass modeling for the Main Belt {asteroids} used in INPOP. Starting from such a list of asteroids, we have been able to remove more than 100 of them without significantly degrading the postfit residuals {and with a significant improvement of the uncertainties} . Moreover, we verified that the postfit masses found with the new modeling (INPOP25c) are consistent with respect to mass estimation found with an independent and different approach, by means of albedo properties of the asteroids. Finally, we proposed a comprehensive comparison with mass estimations already provided in the literature, using different statistical tools. In the end, using the Wasserstein barycenter we merged together different {asteroid mass estimations}, showing how this innovative approach might help in specific cases for masses poorly determined. In the future the technique proposed by boosting decision trees can be of help for improving dynamical modeling within the solar system, such as for TNOs, although computational issues should be addressed in advance and tackled.} 

%\section{Conclusions}

%\begin{acknowledgments}
%\nolinenumbers
\paragraph{\bf Acknowledgments}: This work was supported by the French government through the France 2030 investment plan managed by the National Research Agency (ANR), as part of the Initiative of Excellence Université Côte d’Azur under reference number ANR-15-IDEX-01. The authors are grateful to the Université Côte d’Azur’s Center for High-Performance Computing (OPAL infrastructure) for providing resources and support. VM thanks Alberto Perrella for the fruitful discussions about optimal transport theory. 
%\end{acknowledgments}

\bibliography{Asteroid_selection_bib}{}
\bibliographystyle{aasjournal}

\appendix
\label{sec:Appendix1}

%\section{Complementary material}
\scriptsize
%\input{Table1_mpmc}
%\begin{table}[ht]
%\centering
%\begin{tabular}{cccccc}
%  \hline
%MPC\_number & Mass\_INPOP25c\_kg & Sig\_INPOP25c & Sigma\_o\_Mass\_INPOP25c & Mean\_Wass\_Bary & Sigma\_Wass\_Bary \\ 
%  \hline

\begin{longtable}{c|c|c|c|c|c}%\label{tab:appendix1}
\caption{{List of the asteroid masses obtained in INPOP25c. In the first, second and third column we show respectively the postfit mass, $1\sigma$ uncertainty and noise-to-signal ration found with INPOP25c adjustment to observations. In the fourth and fifth column we present respectively the mean and standard deviation of the Wasserstein barycenters obtained as described in Sec. \ref{sec:other_comp}.}   \label{tab:appendix1}}
 \\ \hline
MPC n. & $m_{\text{25c}}$~[kg] & $\sigma_{\text{25c}}$~[kg] & $\frac{\sigma_{\text{25c}}}{m_{\text{25c}}}$ & $\overline{m}_{\text{Wass.}}$~[kg] & $\sigma_{\text{Wass.}}$ ~[kg] \\ 
  \hline

1 & $9.39 \times 10^{20}$ & $1.39 \times 10^{18}$ & 0.0015 & $9.79 \times 10^{20}$ & $5.82 \times 10^{20}$ \\ 
  2 & $2.13 \times 10^{20}$ & $8.52 \times 10^{17}$ & 0.0040 & $2.37 \times 10^{20}$ & $8.70 \times 10^{19}$ \\ 
  3 & $2.68 \times 10^{19}$ & $4.05 \times 10^{17}$ & 0.0151 & $3.83 \times 10^{19}$ & $3.00 \times 10^{19}$ \\ 
  4 & $2.59 \times 10^{20}$ & $3.95 \times 10^{17}$ & 0.0015 & $2.43 \times 10^{20}$ & $4.91 \times 10^{19}$ \\ 
  5 & $2.65 \times 10^{18}$ & $2.29 \times 10^{17}$ & 0.0865 & $4.30 \times 10^{18}$ & $6.26 \times 10^{18}$ \\ 
  6 & $1.36 \times 10^{19}$ & $5.31 \times 10^{17}$ & 0.0390 & $5.00 \times 10^{19}$ & $7.11 \times 10^{19}$ \\ 
  7 & $1.37 \times 10^{19}$ & $3.32 \times 10^{17}$ & 0.0242 & $3.36 \times 10^{19}$ & $4.25 \times 10^{19}$ \\ 
  8 & $3.64 \times 10^{18}$ & $2.10 \times 10^{17}$ & 0.0577 & $3.92 \times 10^{18}$ & $1.69 \times 10^{18}$ \\ 
  9 & $7.57 \times 10^{18}$ & $4.12 \times 10^{17}$ & 0.0543 & $1.14 \times 10^{19}$ & $1.83 \times 10^{19}$ \\ 
  10 & $8.48 \times 10^{19}$ & $1.87 \times 10^{18}$ & 0.0220 & $8.51 \times 10^{19}$ & $2.11 \times 10^{19}$ \\ 
  11 & $6.36 \times 10^{18}$ & $5.56 \times 10^{17}$ & 0.0874 & $6.01 \times 10^{18}$ & $2.68 \times 10^{18}$ \\ 
  12 & $2.17 \times 10^{18}$ & $2.09 \times 10^{17}$ & 0.0963 & $3.57 \times 10^{18}$ & $3.01 \times 10^{18}$ \\ 
  13 & $4.59 \times 10^{18}$ & $8.06 \times 10^{17}$ & 0.1754 & $5.54 \times 10^{18}$ & $1.77 \times 10^{18}$ \\ 
  14 & $4.34 \times 10^{18}$ & $4.15 \times 10^{17}$ & 0.0955 & $7.91 \times 10^{18}$ & $7.67 \times 10^{18}$ \\ 
  15 & $2.76 \times 10^{19}$ & $8.26 \times 10^{17}$ & 0.0300 & $2.60 \times 10^{19}$ & $5.31 \times 10^{18}$ \\ 
  16 & $2.67 \times 10^{19}$ & $1.19 \times 10^{18}$ & 0.0446 & $2.56 \times 10^{19}$ & $1.54 \times 10^{19}$ \\ 
  17 & $6.97 \times 10^{17}$ & $3.07 \times 10^{17}$ & 0.4414 & $1.15 \times 10^{18}$ & $9.63 \times 10^{17}$ \\ 
  18 & $4.03 \times 10^{18}$ & $2.17 \times 10^{17}$ & 0.0539 & $9.86 \times 10^{18}$ & $1.24 \times 10^{19}$ \\ 
  19 & $7.61 \times 10^{18}$ & $2.48 \times 10^{17}$ & 0.0326 & $7.77 \times 10^{18}$ & $2.65 \times 10^{18}$ \\ 
  20 & $1.69 \times 10^{18}$ & $5.47 \times 10^{17}$ & 0.3238 & $3.80 \times 10^{18}$ & $1.41 \times 10^{18}$ \\ 
  21 & $1.75 \times 10^{18}$ & $4.69 \times 10^{17}$ & 0.2676 & $3.07 \times 10^{18}$ & $2.24 \times 10^{18}$ \\ 
  22 & $8.01 \times 10^{18}$ & $1.17 \times 10^{18}$ & 0.1466 & $8.96 \times 10^{18}$ & $6.42 \times 10^{18}$ \\ 
  23 & $1.52 \times 10^{18}$ & $2.68 \times 10^{17}$ & 0.1764 & $4.42 \times 10^{18}$ & $5.40 \times 10^{18}$ \\ 
  24 & $4.49 \times 10^{18}$ & $1.43 \times 10^{18}$ & 0.3178 & $6.52 \times 10^{18}$ & $2.56 \times 10^{18}$ \\ 
  25 & $1.11 \times 10^{18}$ & $3.96 \times 10^{17}$ & 0.3557 & $7.38 \times 10^{17}$ & $1.74 \times 10^{17}$ \\ 
  26 & $9.51 \times 10^{17}$ & $4.11 \times 10^{17}$ & 0.4317 & $2.74 \times 10^{18}$ & $3.80 \times 10^{18}$ \\ 
  27 & $2.27 \times 10^{18}$ & $3.86 \times 10^{17}$ & 0.1701 & $2.25 \times 10^{18}$ & $6.35 \times 10^{17}$ \\ 
  28 & $1.94 \times 10^{18}$ & $6.50 \times 10^{17}$ & 0.3354 & $5.26 \times 10^{18}$ & $3.50 \times 10^{18}$ \\ 
  29 & $1.60 \times 10^{19}$ & $7.86 \times 10^{17}$ & 0.0490 & $3.06 \times 10^{19}$ & $4.43 \times 10^{19}$ \\ 
  30 & $1.30 \times 10^{18}$ & $4.13 \times 10^{17}$ & 0.3184 & $3.27 \times 10^{18}$ & $2.39 \times 10^{18}$ \\ 
  31 & $1.07 \times 10^{19}$ & $2.32 \times 10^{18}$ & 0.2176 & $1.56 \times 10^{19}$ & $5.51 \times 10^{18}$ \\ 
  32 & $6.45 \times 10^{17}$ & $2.83 \times 10^{17}$ & 0.4381 & $1.59 \times 10^{18}$ & $2.03 \times 10^{18}$ \\ 
  34 & $1.78 \times 10^{18}$ & $5.93 \times 10^{17}$ & 0.3331 & $2.07 \times 10^{18}$ & $4.99 \times 10^{17}$ \\ 
  35 & $6.22 \times 10^{17}$ & $2.23 \times 10^{17}$ & 0.3593 & $1.14 \times 10^{18}$ & $6.97 \times 10^{17}$ \\ 
  36 & $1.52 \times 10^{18}$ & $4.35 \times 10^{17}$ & 0.2862 & $1.85 \times 10^{18}$ & $6.80 \times 10^{17}$ \\ 
  37 & $1.38 \times 10^{18}$ & $4.97 \times 10^{17}$ & 0.3591 & $3.70 \times 10^{18}$ & $4.03 \times 10^{18}$ \\ 
  38 & $2.02 \times 10^{17}$ & $1.00 \times 10^{17}$ & 0.4969 & $1.11 \times 10^{18}$ & $6.22 \times 10^{17}$ \\ 
  39 & $6.56 \times 10^{18}$ & $1.46 \times 10^{18}$ & 0.2224 & $1.21 \times 10^{19}$ & $1.33 \times 10^{19}$ \\ 
  40 & $2.54 \times 10^{18}$ & $4.39 \times 10^{17}$ & 0.1728 & $2.59 \times 10^{18}$ & $8.79 \times 10^{17}$ \\ 
  41 & $7.92 \times 10^{18}$ & $4.49 \times 10^{17}$ & 0.0567 & $7.29 \times 10^{18}$ & $5.17 \times 10^{18}$ \\ 
  42 & $2.33 \times 10^{18}$ & $2.54 \times 10^{17}$ & 0.1090 & $2.40 \times 10^{18}$ & $1.85 \times 10^{18}$ \\ 
  43 & $2.12 \times 10^{17}$ & $8.66 \times 10^{16}$ & 0.4081 & $6.82 \times 10^{17}$ & $5.68 \times 10^{17}$ \\ 
  44 & $2.72 \times 10^{17}$ & $1.34 \times 10^{17}$ & 0.4938 & $2.41 \times 10^{18}$ & $3.55 \times 10^{18}$ \\ 
  45 & $6.35 \times 10^{18}$ & $1.01 \times 10^{18}$ & 0.1588 & $6.27 \times 10^{18}$ & $2.01 \times 10^{18}$ \\ 
  46 & $5.24 \times 10^{17}$ & $2.34 \times 10^{17}$ & 0.4457 & $1.03 \times 10^{18}$ & $5.34 \times 10^{17}$ \\ 
  47 & $6.00 \times 10^{17}$ & $2.91 \times 10^{17}$ & 0.4856 & $1.63 \times 10^{18}$ & $9.71 \times 10^{17}$ \\ 
  48 & $9.11 \times 10^{18}$ & $2.84 \times 10^{18}$ & 0.3118 & $1.02 \times 10^{19}$ & $4.84 \times 10^{18}$ \\ 
  49 & $2.51 \times 10^{18}$ & $8.91 \times 10^{17}$ & 0.3552 & $3.46 \times 10^{18}$ & $1.83 \times 10^{18}$ \\ 
  50 & $2.85 \times 10^{17}$ & $1.27 \times 10^{17}$ & 0.4455 & $5.66 \times 10^{17}$ & $2.66 \times 10^{17}$ \\ 
  51 & $2.68 \times 10^{18}$ & $5.43 \times 10^{17}$ & 0.2026 & $2.76 \times 10^{18}$ & $1.42 \times 10^{18}$ \\ 
  52 & $3.11 \times 10^{19}$ & $1.90 \times 10^{18}$ & 0.0611 & $2.86 \times 10^{19}$ & $1.10 \times 10^{19}$ \\ 
  53 & $6.02 \times 10^{17}$ & $2.70 \times 10^{17}$ & 0.4488 & $8.47 \times 10^{17}$ & $3.51 \times 10^{17}$ \\ 
  54 & $2.79 \times 10^{18}$ & $6.95 \times 10^{17}$ & 0.2493 & $3.77 \times 10^{18}$ & $2.57 \times 10^{18}$ \\ 
  56 & $3.26 \times 10^{18}$ & $3.94 \times 10^{17}$ & 0.1208 & $2.99 \times 10^{18}$ & $7.89 \times 10^{17}$ \\ 
  57 & $3.60 \times 10^{18}$ & $1.34 \times 10^{18}$ & 0.3717 & $8.07 \times 10^{18}$ & $5.63 \times 10^{18}$ \\ 
  58 & $1.87 \times 10^{17}$ & $9.33 \times 10^{16}$ & 0.4980 & $3.71 \times 10^{17}$ & $2.64 \times 10^{17}$ \\ 
  59 & $3.56 \times 10^{18}$ & $9.40 \times 10^{17}$ & 0.2642 & $3.09 \times 10^{18}$ & $1.46 \times 10^{18}$ \\ 
  60 & $1.98 \times 10^{17}$ & $8.64 \times 10^{16}$ & 0.4363 & $4.95 \times 10^{17}$ & $4.08 \times 10^{17}$ \\ 
  62 & $2.60 \times 10^{17}$ & $1.29 \times 10^{17}$ & 0.4983 & $5.70 \times 10^{17}$ & $3.19 \times 10^{17}$ \\ 
  63 & $8.29 \times 10^{17}$ & $3.31 \times 10^{17}$ & 0.4000 & $9.71 \times 10^{17}$ & $2.37 \times 10^{17}$ \\ 
  65 & $9.08 \times 10^{18}$ & $1.77 \times 10^{18}$ & 0.1952 & $1.51 \times 10^{19}$ & $6.13 \times 10^{18}$ \\ 
  68 & $1.90 \times 10^{18}$ & $6.88 \times 10^{17}$ & 0.3626 & $5.13 \times 10^{18}$ & $5.44 \times 10^{18}$ \\ 
  69 & $3.87 \times 10^{18}$ & $8.83 \times 10^{17}$ & 0.2279 & $7.71 \times 10^{18}$ & $7.44 \times 10^{18}$ \\ 
  70 & $1.47 \times 10^{18}$ & $4.59 \times 10^{17}$ & 0.3114 & $2.11 \times 10^{18}$ & $7.77 \times 10^{17}$ \\ 
  71 & $1.01 \times 10^{18}$ & $4.59 \times 10^{17}$ & 0.4548 & $2.25 \times 10^{18}$ & $5.79 \times 10^{18}$ \\ 
  72 & $9.50 \times 10^{17}$ & $3.22 \times 10^{17}$ & 0.3390 & $5.50 \times 10^{17}$ & $1.43 \times 10^{17}$ \\ 
  74 & $2.24 \times 10^{18}$ & $6.10 \times 10^{17}$ & 0.2722 & $1.52 \times 10^{18}$ & $4.94 \times 10^{17}$ \\ 
  75 & $4.81 \times 10^{17}$ & $1.90 \times 10^{17}$ & 0.3960 & $6.35 \times 10^{17}$ & $6.40 \times 10^{17}$ \\ 
  76 & $2.99 \times 10^{18}$ & $1.04 \times 10^{18}$ & 0.3466 & $3.69 \times 10^{18}$ & $1.32 \times 10^{18}$ \\ 
  77 & $5.36 \times 10^{17}$ & $2.58 \times 10^{17}$ & 0.4816 & $8.92 \times 10^{17}$ & $8.05 \times 10^{17}$ \\ 
  78 & $3.36 \times 10^{18}$ & $3.97 \times 10^{17}$ & 0.1181 & $2.29 \times 10^{18}$ & $1.67 \times 10^{18}$ \\ 
  79 & $5.54 \times 10^{17}$ & $2.27 \times 10^{17}$ & 0.4104 & $8.24 \times 10^{17}$ & $9.61 \times 10^{17}$ \\ 
  80 & $6.34 \times 10^{17}$ & $2.23 \times 10^{17}$ & 0.3515 & $7.15 \times 10^{17}$ & $4.17 \times 10^{17}$ \\ 
  81 & $1.27 \times 10^{18}$ & $5.17 \times 10^{17}$ & 0.4060 & $1.90 \times 10^{18}$ & $5.69 \times 10^{17}$ \\ 
  82 & $5.02 \times 10^{17}$ & $2.36 \times 10^{17}$ & 0.4704 & $1.00 \times 10^{18}$ & $7.91 \times 10^{17}$ \\ 
  83 & $3.87 \times 10^{17}$ & $1.86 \times 10^{17}$ & 0.4798 & $5.76 \times 10^{17}$ & $5.90 \times 10^{17}$ \\ 
  84 & $4.17 \times 10^{17}$ & $1.26 \times 10^{17}$ & 0.3021 & $4.92 \times 10^{17}$ & $2.58 \times 10^{17}$ \\ 
  85 & $8.31 \times 10^{18}$ & $6.43 \times 10^{17}$ & 0.0774 & $4.05 \times 10^{18}$ & $6.39 \times 10^{17}$ \\ 
  86 & $1.93 \times 10^{18}$ & $8.40 \times 10^{17}$ & 0.4348 & $1.68 \times 10^{18}$ & $6.25 \times 10^{17}$ \\ 
  87 & $1.83 \times 10^{19}$ & $2.35 \times 10^{18}$ & 0.1286 & $1.81 \times 10^{19}$ & $6.46 \times 10^{18}$ \\ 
  88 & $7.18 \times 10^{18}$ & $2.08 \times 10^{16}$ & 0.0029 & $7.50 \times 10^{18}$ & $2.38 \times 10^{18}$ \\ 
  89 & $3.81 \times 10^{18}$ & $5.38 \times 10^{17}$ & 0.1411 & $6.46 \times 10^{18}$ & $3.70 \times 10^{18}$ \\ 
  90 & $1.05 \times 10^{18}$ & $5.08 \times 10^{17}$ & 0.4817 & $1.20 \times 10^{18}$ & $4.92 \times 10^{17}$ \\ 
  91 & $6.17 \times 10^{17}$ & $2.87 \times 10^{17}$ & 0.4645 & $6.38 \times 10^{17}$ & $4.12 \times 10^{17}$ \\ 
  92 & $2.14 \times 10^{18}$ & $9.72 \times 10^{17}$ & 0.4533 & $9.38 \times 10^{18}$ & $6.77 \times 10^{18}$ \\ 
  93 & $2.14 \times 10^{18}$ & $6.84 \times 10^{17}$ & 0.3194 & $2.65 \times 10^{18}$ & $1.33 \times 10^{18}$ \\ 
  94 & $5.08 \times 10^{18}$ & $1.67 \times 10^{18}$ & 0.3299 & $5.42 \times 10^{18}$ & $2.76 \times 10^{18}$ \\ 
  95 & $3.78 \times 10^{17}$ & $1.89 \times 10^{17}$ & 0.4993 & $1.88 \times 10^{18}$ & $1.17 \times 10^{18}$ \\ 
  96 & $2.90 \times 10^{18}$ & $1.05 \times 10^{18}$ & 0.3622 & $4.88 \times 10^{18}$ & $1.92 \times 10^{18}$ \\ 
  97 & $7.45 \times 10^{17}$ & $2.93 \times 10^{17}$ & 0.3930 & $1.91 \times 10^{18}$ & $1.50 \times 10^{18}$ \\ 
  98 & $9.48 \times 10^{16}$ & $4.72 \times 10^{16}$ & 0.4979 & $7.57 \times 10^{17}$ & $7.98 \times 10^{17}$ \\ 
  99 & $6.06 \times 10^{17}$ & $2.37 \times 10^{17}$ & 0.3905 & $4.70 \times 10^{17}$ & $1.58 \times 10^{17}$ \\ 
  100 & $6.70 \times 10^{17}$ & $3.28 \times 10^{17}$ & 0.4897 & $2.32 \times 10^{18}$ & $1.48 \times 10^{18}$ \\ 
  102 & $1.83 \times 10^{18}$ & $4.26 \times 10^{17}$ & 0.2324 & $1.27 \times 10^{18}$ & $3.23 \times 10^{17}$ \\ 
  103 & $1.23 \times 10^{18}$ & $5.59 \times 10^{17}$ & 0.4550 & $2.75 \times 10^{18}$ & $2.40 \times 10^{18}$ \\ 
  104 & $1.22 \times 10^{18}$ & $7.81 \times 10^{17}$ & 0.6398 & $1.53 \times 10^{18}$ & $5.40 \times 10^{17}$ \\ 
  105 & $1.14 \times 10^{18}$ & $3.70 \times 10^{17}$ & 0.3239 & $1.36 \times 10^{18}$ & $2.50 \times 10^{18}$ \\ 
  106 & $1.91 \times 10^{18}$ & $1.01 \times 10^{18}$ & 0.5279 & $3.92 \times 10^{18}$ & $1.84 \times 10^{18}$ \\ 
  107 & $1.44 \times 10^{19}$ & $2.31 \times 10^{18}$ & 0.1604 & $1.43 \times 10^{19}$ & $4.62 \times 10^{18}$ \\ 
  109 & $4.10 \times 10^{17}$ & $1.77 \times 10^{17}$ & 0.4310 & $5.00 \times 10^{17}$ & $3.29 \times 10^{17}$ \\ 
  110 & $1.24 \times 10^{18}$ & $6.41 \times 10^{17}$ & 0.5178 & $2.29 \times 10^{18}$ & $1.09 \times 10^{18}$ \\ 
  111 & $1.03 \times 10^{18}$ & $4.73 \times 10^{17}$ & 0.4580 & $1.40 \times 10^{18}$ & $9.26 \times 10^{17}$ \\ 
  113 & $2.64 \times 10^{17}$ & $1.28 \times 10^{17}$ & 0.4859 & $3.37 \times 10^{17}$ & $3.82 \times 10^{17}$ \\ 
  114 & $8.11 \times 10^{17}$ & $3.50 \times 10^{17}$ & 0.4319 & $1.17 \times 10^{18}$ & $7.47 \times 10^{17}$ \\ 
  115 & $2.99 \times 10^{17}$ & $1.44 \times 10^{17}$ & 0.4836 & $1.15 \times 10^{18}$ & $1.36 \times 10^{18}$ \\ 
  117 & $6.93 \times 10^{18}$ & $1.84 \times 10^{18}$ & 0.2650 & $5.43 \times 10^{18}$ & $1.33 \times 10^{18}$ \\ 
  118 & $1.49 \times 10^{17}$ & $7.34 \times 10^{16}$ & 0.4933 & $1.75 \times 10^{17}$ & $1.81 \times 10^{17}$ \\ 
  120 & $4.03 \times 10^{18}$ & $1.50 \times 10^{18}$ & 0.3727 & $5.01 \times 10^{18}$ & $2.20 \times 10^{18}$ \\ 
  121 & $4.50 \times 10^{18}$ & $1.42 \times 10^{18}$ & 0.3156 & $7.02 \times 10^{18}$ & $3.55 \times 10^{18}$ \\ 
  127 & $6.25 \times 10^{17}$ & $3.04 \times 10^{17}$ & 0.4861 & $1.11 \times 10^{18}$ & $6.07 \times 10^{17}$ \\ 
  128 & $3.23 \times 10^{18}$ & $8.72 \times 10^{17}$ & 0.2703 & $4.54 \times 10^{18}$ & $1.59 \times 10^{18}$ \\ 
  129 & $4.04 \times 10^{18}$ & $7.36 \times 10^{17}$ & 0.1821 & $9.20 \times 10^{18}$ & $2.44 \times 10^{19}$ \\ 
  130 & $7.78 \times 10^{18}$ & $1.48 \times 10^{18}$ & 0.1906 & $8.41 \times 10^{18}$ & $3.38 \times 10^{18}$ \\ 
  134 & $1.98 \times 10^{18}$ & $6.23 \times 10^{17}$ & 0.3144 & $1.59 \times 10^{18}$ & $1.61 \times 10^{18}$ \\ 
  135 & $9.28 \times 10^{17}$ & $3.55 \times 10^{17}$ & 0.3821 & $8.32 \times 10^{17}$ & $1.72 \times 10^{17}$ \\ 
  137 & $2.18 \times 10^{18}$ & $8.14 \times 10^{17}$ & 0.3729 & $3.16 \times 10^{18}$ & $1.07 \times 10^{18}$ \\ 
  139 & $1.53 \times 10^{18}$ & $4.55 \times 10^{17}$ & 0.2975 & $3.64 \times 10^{18}$ & $1.67 \times 10^{18}$ \\ 
  140 & $2.47 \times 10^{18}$ & $5.07 \times 10^{17}$ & 0.2055 & $2.09 \times 10^{18}$ & $4.61 \times 10^{17}$ \\ 
  141 & $1.73 \times 10^{18}$ & $4.59 \times 10^{17}$ & 0.2654 & $1.28 \times 10^{18}$ & $4.43 \times 10^{17}$ \\ 
  148 & $7.87 \times 10^{17}$ & $3.78 \times 10^{17}$ & 0.4808 & $9.10 \times 10^{17}$ & $3.96 \times 10^{17}$ \\ 
  150 & $2.14 \times 10^{18}$ & $3.78 \times 10^{17}$ & 0.1760 & $1.93 \times 10^{18}$ & $1.03 \times 10^{18}$ \\ 
  154 & $6.43 \times 10^{18}$ & $1.80 \times 10^{18}$ & 0.2796 & $7.03 \times 10^{18}$ & $3.54 \times 10^{18}$ \\ 
  159 & $1.96 \times 10^{18}$ & $9.39 \times 10^{17}$ & 0.4795 & $2.49 \times 10^{18}$ & $7.44 \times 10^{17}$ \\ 
  160 & $4.28 \times 10^{17}$ & $2.10 \times 10^{17}$ & 0.4920 & $5.60 \times 10^{17}$ & $4.28 \times 10^{17}$ \\ 
  162 & $1.90 \times 10^{18}$ & $6.78 \times 10^{17}$ & 0.3559 & $1.42 \times 10^{18}$ & $4.26 \times 10^{17}$ \\ 
  163 & $2.28 \times 10^{16}$ & $1.14 \times 10^{16}$ & 0.4993 & $3.16 \times 10^{17}$ & $1.47 \times 10^{17}$ \\ 
  164 & $2.22 \times 10^{18}$ & $3.25 \times 10^{17}$ & 0.1463 & $1.78 \times 10^{18}$ & $5.28 \times 10^{17}$ \\ 
  165 & $7.36 \times 10^{17}$ & $3.63 \times 10^{17}$ & 0.4941 & $4.79 \times 10^{18}$ & $3.03 \times 10^{18}$ \\ 
  172 & $4.83 \times 10^{17}$ & $2.24 \times 10^{17}$ & 0.4643 & $4.92 \times 10^{17}$ & $2.18 \times 10^{17}$ \\ 
  173 & $7.80 \times 10^{17}$ & $3.59 \times 10^{17}$ & 0.4599 & $1.29 \times 10^{18}$ & $4.55 \times 10^{17}$ \\ 
  187 & $2.02 \times 10^{18}$ & $3.01 \times 10^{17}$ & 0.1485 & $1.99 \times 10^{18}$ & $1.11 \times 10^{18}$ \\ 
  191 & $3.50 \times 10^{17}$ & $1.74 \times 10^{17}$ & 0.4971 & $7.25 \times 10^{17}$ & $3.41 \times 10^{17}$ \\ 
  192 & $1.62 \times 10^{18}$ & $3.61 \times 10^{17}$ & 0.2234 & $3.03 \times 10^{18}$ & $2.64 \times 10^{18}$ \\ 
  194 & $4.29 \times 10^{18}$ & $5.07 \times 10^{17}$ & 0.1181 & $3.81 \times 10^{18}$ & $1.73 \times 10^{18}$ \\ 
  200 & $2.03 \times 10^{18}$ & $8.16 \times 10^{17}$ & 0.4015 & $2.25 \times 10^{18}$ & $8.50 \times 10^{17}$ \\ 
  201 & $1.28 \times 10^{18}$ & $5.45 \times 10^{17}$ & 0.4266 & $1.31 \times 10^{18}$ & $9.32 \times 10^{17}$ \\ 
  209 & $1.37 \times 10^{18}$ & $6.77 \times 10^{17}$ & 0.4947 & $1.96 \times 10^{18}$ & $1.01 \times 10^{18}$ \\ 
  211 & $2.08 \times 10^{18}$ & $9.16 \times 10^{17}$ & 0.4403 & $2.71 \times 10^{18}$ & $8.81 \times 10^{17}$ \\ 
  212 & $4.36 \times 10^{18}$ & $1.55 \times 10^{18}$ & 0.3553 & $2.95 \times 10^{18}$ & $1.23 \times 10^{18}$ \\ 
  216 & $4.72 \times 10^{18}$ & $7.78 \times 10^{17}$ & 0.1648 & $6.62 \times 10^{18}$ & $7.83 \times 10^{18}$ \\ 
  221 & $1.18 \times 10^{18}$ & $4.53 \times 10^{17}$ & 0.3851 & $1.38 \times 10^{18}$ & $4.06 \times 10^{17}$ \\ 
  227 & $1.72 \times 10^{18}$ & $7.32 \times 10^{17}$ & 0.4242 & $1.45 \times 10^{18}$ & $4.49 \times 10^{17}$ \\ 
  230 & $1.98 \times 10^{18}$ & $6.12 \times 10^{17}$ & 0.3089 & $2.48 \times 10^{18}$ & $1.71 \times 10^{18}$ \\ 
  238 & $1.15 \times 10^{18}$ & $5.34 \times 10^{17}$ & 0.4642 & $2.64 \times 10^{18}$ & $1.18 \times 10^{18}$ \\ 
  241 & $5.74 \times 10^{18}$ & $1.63 \times 10^{18}$ & 0.2839 & $6.08 \times 10^{18}$ & $2.23 \times 10^{18}$ \\ 
  247 & $2.96 \times 10^{17}$ & $1.46 \times 10^{17}$ & 0.4945 & $6.19 \times 10^{17}$ & $5.59 \times 10^{17}$ \\ 
  250 & $1.97 \times 10^{18}$ & $9.14 \times 10^{17}$ & 0.4632 & $3.70 \times 10^{18}$ & $2.62 \times 10^{18}$ \\ 
  259 & $1.35 \times 10^{18}$ & $6.20 \times 10^{17}$ & 0.4586 & $5.44 \times 10^{18}$ & $3.07 \times 10^{18}$ \\ 
  266 & $2.00 \times 10^{18}$ & $8.59 \times 10^{17}$ & 0.4301 & $1.46 \times 10^{18}$ & $5.41 \times 10^{17}$ \\ 
  268 & $4.97 \times 10^{16}$ & $2.48 \times 10^{16}$ & 0.4999 & $1.34 \times 10^{18}$ & $6.33 \times 10^{17}$ \\ 
  283 & $3.01 \times 10^{18}$ & $9.89 \times 10^{17}$ & 0.3284 & $2.19 \times 10^{18}$ & $6.81 \times 10^{17}$ \\ 
  287 & $1.79 \times 10^{17}$ & $8.88 \times 10^{16}$ & 0.4953 & $4.51 \times 10^{17}$ & $4.24 \times 10^{17}$ \\ 
  308 & $2.08 \times 10^{18}$ & $8.50 \times 10^{17}$ & 0.4089 & $3.32 \times 10^{18}$ & $1.24 \times 10^{18}$ \\ 
  313 & $5.83 \times 10^{17}$ & $1.59 \times 10^{17}$ & 0.2724 & $6.25 \times 10^{17}$ & $2.43 \times 10^{17}$ \\ 
  324 & $1.06 \times 10^{19}$ & $1.87 \times 10^{17}$ & 0.0177 & $1.04 \times 10^{19}$ & $4.87 \times 10^{18}$ \\ 
  326 & $7.20 \times 10^{17}$ & $2.03 \times 10^{17}$ & 0.2822 & $4.88 \times 10^{17}$ & $3.62 \times 10^{17}$ \\ 
  328 & $2.10 \times 10^{18}$ & $9.14 \times 10^{17}$ & 0.4356 & $2.18 \times 10^{18}$ & $6.79 \times 10^{17}$ \\ 
  336 & $1.13 \times 10^{17}$ & $5.55 \times 10^{16}$ & 0.4924 & $1.96 \times 10^{17}$ & $7.07 \times 10^{16}$ \\ 
  337 & $5.94 \times 10^{17}$ & $2.36 \times 10^{17}$ & 0.3969 & $4.94 \times 10^{17}$ & $2.11 \times 10^{17}$ \\ 
  345 & $6.53 \times 10^{17}$ & $2.71 \times 10^{17}$ & 0.4155 & $7.46 \times 10^{17}$ & $3.37 \times 10^{17}$ \\ 
  346 & $9.07 \times 10^{17}$ & $4.27 \times 10^{17}$ & 0.4712 & $3.81 \times 10^{18}$ & $5.26 \times 10^{18}$ \\ 
  354 & $8.26 \times 10^{18}$ & $7.68 \times 10^{17}$ & 0.0929 & $1.68 \times 10^{19}$ & $9.30 \times 10^{18}$ \\ 
  356 & $1.21 \times 10^{18}$ & $3.92 \times 10^{17}$ & 0.3244 & $1.52 \times 10^{18}$ & $7.91 \times 10^{17}$ \\ 
  362 & $5.70 \times 10^{17}$ & $2.80 \times 10^{17}$ & 0.4915 & $5.49 \times 10^{17}$ & $2.28 \times 10^{17}$ \\ 
  372 & $9.95 \times 10^{18}$ & $1.91 \times 10^{18}$ & 0.1918 & $8.42 \times 10^{18}$ & $2.20 \times 10^{18}$ \\ 
  386 & $3.92 \times 10^{18}$ & $9.74 \times 10^{17}$ & 0.2483 & $6.00 \times 10^{18}$ & $3.23 \times 10^{18}$ \\ 
  387 & $2.09 \times 10^{18}$ & $5.52 \times 10^{17}$ & 0.2645 & $3.93 \times 10^{18}$ & $3.05 \times 10^{18}$ \\ 
  393 & $1.56 \times 10^{18}$ & $4.22 \times 10^{17}$ & 0.2712 & $2.08 \times 10^{18}$ & $1.18 \times 10^{18}$ \\ 
  404 & $5.30 \times 10^{17}$ & $2.17 \times 10^{17}$ & 0.4089 & $7.46 \times 10^{17}$ & $3.13 \times 10^{17}$ \\ 
  405 & $1.34 \times 10^{18}$ & $2.34 \times 10^{17}$ & 0.1740 & $1.38 \times 10^{18}$ & $4.95 \times 10^{17}$ \\ 
  407 & $6.49 \times 10^{17}$ & $3.31 \times 10^{17}$ & 0.5097 & $6.79 \times 10^{17}$ & $3.19 \times 10^{17}$ \\ 
  409 & $5.45 \times 10^{18}$ & $8.17 \times 10^{17}$ & 0.1499 & $7.48 \times 10^{18}$ & $3.62 \times 10^{18}$ \\ 
  410 & $2.37 \times 10^{18}$ & $4.86 \times 10^{17}$ & 0.2049 & $2.85 \times 10^{18}$ & $9.71 \times 10^{17}$ \\ 
  416 & $7.38 \times 10^{17}$ & $3.39 \times 10^{17}$ & 0.4591 & $1.74 \times 10^{18}$ & $1.79 \times 10^{18}$ \\ 
  419 & $2.88 \times 10^{18}$ & $3.50 \times 10^{17}$ & 0.1214 & $1.99 \times 10^{18}$ & $9.83 \times 10^{17}$ \\ 
  431 & $7.83 \times 10^{17}$ & $3.81 \times 10^{17}$ & 0.4859 & $8.01 \times 10^{17}$ & $2.78 \times 10^{17}$ \\ 
  444 & $4.28 \times 10^{18}$ & $1.02 \times 10^{18}$ & 0.2377 & $3.69 \times 10^{18}$ & $2.28 \times 10^{18}$ \\ 
  451 & $1.30 \times 10^{19}$ & $2.94 \times 10^{18}$ & 0.2254 & $1.44 \times 10^{19}$ & $5.48 \times 10^{18}$ \\ 
  469 & $3.05 \times 10^{18}$ & $8.96 \times 10^{17}$ & 0.2941 & $2.87 \times 10^{18}$ & $9.52 \times 10^{17}$ \\ 
  481 & $4.23 \times 10^{17}$ & $1.90 \times 10^{17}$ & 0.4497 & $9.36 \times 10^{17}$ & $5.27 \times 10^{17}$ \\ 
  511 & $4.88 \times 10^{19}$ & $2.16 \times 10^{18}$ & 0.0442 & $3.91 \times 10^{19}$ & $1.12 \times 10^{19}$ \\ 
  516 & $5.33 \times 10^{17}$ & $2.00 \times 10^{17}$ & 0.3745 & $1.01 \times 10^{18}$ & $1.10 \times 10^{18}$ \\ 
  532 & $8.46 \times 10^{18}$ & $6.32 \times 10^{17}$ & 0.0747 & $1.94 \times 10^{19}$ & $2.47 \times 10^{19}$ \\ 
  654 & $9.38 \times 10^{17}$ & $1.65 \times 10^{17}$ & 0.1756 & $1.20 \times 10^{18}$ & $1.05 \times 10^{18}$ \\ 
  674 & $1.08 \times 10^{18}$ & $4.63 \times 10^{17}$ & 0.4280 & $2.31 \times 10^{18}$ & $3.25 \times 10^{18}$ \\ 
  690 & $5.91 \times 10^{18}$ & $1.47 \times 10^{18}$ & 0.2485 & $3.66 \times 10^{18}$ & $1.27 \times 10^{18}$ \\ 
  694 & $8.38 \times 10^{17}$ & $2.10 \times 10^{17}$ & 0.2513 & $7.99 \times 10^{17}$ & $2.61 \times 10^{17}$ \\ 
  696 & $5.72 \times 10^{17}$ & $2.81 \times 10^{17}$ & 0.4917 & $5.69 \times 10^{17}$ & $1.83 \times 10^{17}$ \\ 
  704 & $3.76 \times 10^{19}$ & $1.81 \times 10^{18}$ & 0.0482 & $3.89 \times 10^{19}$ & $1.25 \times 10^{19}$ \\ 
  712 & $2.74 \times 10^{18}$ & $5.65 \times 10^{17}$ & 0.2060 & $1.89 \times 10^{18}$ & $3.83 \times 10^{17}$ \\ 
  747 & $6.32 \times 10^{18}$ & $5.48 \times 10^{17}$ & 0.0868 & $5.56 \times 10^{18}$ & $2.58 \times 10^{18}$ \\ 
  751 & $5.50 \times 10^{17}$ & $2.48 \times 10^{17}$ & 0.4520 & $8.85 \times 10^{17}$ & $3.90 \times 10^{17}$ \\ 
  760 & $3.82 \times 10^{17}$ & $1.83 \times 10^{17}$ & 0.4806 & $2.07 \times 10^{18}$ & $1.56 \times 10^{18}$ \\ 
  786 & $1.87 \times 10^{18}$ & $7.74 \times 10^{17}$ & 0.4144 & $1.64 \times 10^{18}$ & $5.10 \times 10^{17}$ \\ 
  804 & $1.71 \times 10^{18}$ & $4.08 \times 10^{17}$ & 0.2388 & $2.72 \times 10^{18}$ & $7.18 \times 10^{17}$ \\ 
  1021 & $9.63 \times 10^{17}$ & $3.66 \times 10^{17}$ & 0.3804 & $1.01 \times 10^{18}$ & $4.97 \times 10^{17}$ \\ 
   \hline
\end{longtable}

\end{document}